\documentclass[11pt,letter]{article}
\usepackage{amsmath}
\usepackage{amssymb}
\usepackage[utf8]{inputenc}
\usepackage[margin=1in]{geometry}
\usepackage[hyphens]{url}
\usepackage{hyperref}
\hypersetup{
  breaklinks=true,
  colorlinks   = true, 
  urlcolor     = blue, 
  linkcolor    = blue, 
  citecolor   = blue
}
\usepackage{graphicx}
\usepackage{amsthm}
\usepackage{xspace}
\usepackage{xargs}
\usepackage[conf={restate,no link to proof}]{proof-at-the-end}
\usepackage{comment}
\usepackage{authblk}
\usepackage{lineno}

\linenumbers

\def\notationcolor{black} 

\newcommand{\notation}[2]{\newcommand{#1}{{\textcolor{\notationcolor}{\ensuremath{#2}}}}}

\newcommand{\term}[2]{\newcommand{#1}{\textcolor{\notationcolor}{#2}\xspace}}

\notation{\mech}{M}
\notation{\data}{\mathcal{D}}
\notation{\datarv}{\mathcal{X}}
\notation{\dataspace}{\mathcal{U}}
\notation{\record}{r}
\notation{\outp}{\omega}
\notation{\randalg}{A}
\notation{\datb}{\data^\prime}
\notation{\query}{q}
\notation{\power}{1-\beta}
\notation{\typetwo}{\beta}
\notation{\level}{\ell}
\notation{\prior}{P_\mathfrak{A}}
\notation{\recordrv}{R}
\notation{\datasize}{n}
\notation{\psample}{\textup{psample}} 
\notation{\public}{R}
\notation{\person}{p}
\notation{\sensitive}{s}
\notation{\attackinfo}{A}
\notation{\precord}{r} 
\notation{\svalue}{S}
\newcommand{\yes}{y}
\newcommand{\no}{n}
\newcommand{\maybe}{?}

\newcommandx*\plrv[3][1=\ensuremath{\data_1}, 2=\ensuremath{\data_2}, 3=\mech]{\ensuremath{\mathcal{\textcolor{\notationcolor}{L}}_{#1,#2}^{#3}}\xspace} 

\newcommandx*\rem[1][1=\ensuremath{\data}]{\ensuremath{\textcolor{\notationcolor}{#1^{-}}}} 

\newcommand{\add}[2]{\ensuremath{#2\textcolor{blue}{\cup\{}#1\textcolor{blue}{\}}}}

\newcommandx*\repl[2][1=\ensuremath{\data},2=\record]{\add{#2}{\rem[#1]}}

\term{\pbdp}{pbdp}
\term{\pbdplong}{probabilistically-bounded differential privacy}
\term{\Pbdplong}{Probabilistically-bounded differential privacy}
\term{\target}{Jessie}

\newcommand{\change}[1]{\textcolor{black}{#1}}

\newcommand{\changetwo}[1]{\textcolor{black}{#1}}

\allowdisplaybreaks

\title{An In-Depth Examination of Requirements for Disclosure Risk Assessment}

\author[a]{Ron S. Jarmin}
\author[b]{John M. Abowd}
\author[a]{Robert Ashmead}
\author[a]{Ryan Cumings-Menon}
\author[a]{Nathan Goldschlag}
\author[a]{Michael B. Hawes}
\author[a,c]{Sallie Ann Keller}
\author[a,d]{Daniel Kifer}
\author[a]{Philip Leclerc}
\author[a,e]{Jerome P. Reiter}
\author[a]{Rolando A. Rodríguez}
\author[f]{Ian Schmutte}
\author[a]{Victoria A. Velkoff}
\author[a]{Pavel Zhuravlev}

\affil[a]{U.S. Census Bureau, 4600 Silver Hill Road, Washington, DC 20233}
\affil[b]{Cornell University, Department of Economics, Ithaca, NY 14853}
\affil[c]{University of Virginia, Biocomplexity Institute, Charlottesville, VA 22904}
\affil[d]{Penn State University, Department of Computer Science and Engineering, University Park, PA 16802}
\affil[e]{University of Georgia, Department of Economics, Athens, GA 30602}
\affil[f]{Duke University, Department of Statistical Science, Durham, NC 27708}

\begin{document}
\nolinenumbers

\date{October 13, 2023}
\maketitle

\section*{Acknowledgement and Disclaimer}
The views expressed in this perspective are those of the authors and not the U.S. Census Bureau. A slightly shorter version of this article with more of the material in the main text moved to the appendices is forthcoming as an invited \emph{Proceedings of the National Academy of Sciences} Perspective (\url{https://doi.org/10.1073/pnas.2220558120}). We thank the editor, Ariel Pakes, and referees from that journal for their careful review and comments.

\newpage
\section*{Abstract}
The use of formal privacy to protect the confidentiality of responses in the 2020 Decennial Census of Population and Housing has triggered renewed interest and debate over how to measure the disclosure risks and societal benefits of the published data products. Following long-established precedent in economics and statistics, we argue that any proposal for quantifying disclosure risk should be based on pre-specified, objective criteria. Such criteria should be used to compare  methodologies to identify those with the most desirable properties. We illustrate this approach, using simple desiderata, to evaluate  the absolute disclosure risk framework, the counterfactual framework underlying differential privacy, and prior-to-posterior comparisons. We conclude that satisfying all the desiderata is impossible, but counterfactual comparisons satisfy the most while absolute disclosure risk satisfies the fewest. Furthermore, we explain that  many of the criticisms levied against differential privacy  would be levied against any technology that is not equivalent to direct, unrestricted access to confidential data. Thus, more research is needed, but in the near-term, the counterfactual approach appears best-suited for privacy-utility analysis. 
\\
\\
\noindent \emph{Keywords:} federal statistical system, data disclosure risk, data access

\newpage

\section*{Introduction}

Data products from statistical agencies are vital to the functioning of many countries---supporting the distribution of political representation, business decision-making, research, government planning, and policy-making.
The social welfare derived from these data products depends on their utility and also their confidentiality protections, which are mandated by legal and ethical considerations. \change{By law, these confidentiality protections generally extend to all respondent information, regardless of how sensitive any particular response element may appear to be \cite{baldrige}. While statistical agencies regularly enhance confidentiality protections for obviously sensitive elements, the disclosure of which would likely cause   harm to a respondent (for example, sensitive medical information), in practice statistical agencies' reliance on the voluntary participation of respondents and the information that those particular respondents may or may not consider sensitive is inherently contextual in nature \cite{Nissenbaum+2009}. This means that it would be impractical for agencies to discount the importance of confidentiality even for response information that may not readily be seen as particularly sensitive.}  

Since 1978, the federal statistical system has acknowledged that any public release of data generated from a confidential data source carries some risk of disclosure \cite{FCSM:SWP2:1978}. Balancing data utility and disclosure risk is often accomplished using a disclosure avoidance system, whose input consists of the confidential data and whose output consists of data products with a certain degree of confidentiality protection.

A disclosure avoidance system can apply a variety of statistical disclosure limitation (SDL) techniques to the input data, including suppression, coarsening, noise infusion, and aggregation \changetwo{(see Appendix \ref{sec:si-flexibility} for a comparative discussion)}.
The design of these systems and the tuning of their various parameters (e.g., \change{underlying population thresholds or the amount of noise to be added) }
requires a careful cost-benefit analysis. 
The benefit side depends on \change{stake-holder engagement and consideration of} a wide range of  use-cases. 
Quantifying the cost side requires a disclosure risk assessment methodology. This is a tricky task that is difficult to perform well---many notions of disclosure risk that at first seem intuitive can leave confidential data vulnerable by underestimating risk, or damage the data utility by overestimating risk.
Quantifying disclosure risk is not a  solved problem, but significant progress has been made in the last few decades,  partly driven by systematic approaches that avoid presupposing a specific measure of disclosure risk. Instead, they propose a set of desirable properties and then identify specific disclosure risk measures that satisfy those properties. 

In this paper, we illustrate such an approach in a minimally technical way. We lay out a set of relatively non-controversial desiderata for disclosure risk assessment methods. We use these desiderata to analyze   three major frameworks:  \emph{absolute disclosure risk}, \emph{prior-to-posterior comparisons}, and the \emph{counterfactual comparisons} that motivate differential privacy \cite{dwork:etal:2006,dwork:2006,dpcausal}.
Based on this analysis, we conclude that it is impossible to simultaneously satisfy all the desiderata.
However, differential privacy and its counterfactual framework are supported by the best available science, clearly perform the best and, therefore, should be recommended for use at statistical agencies \change{when SDL implementations based on this framework can be effectively implemented. The potential for effective implementation is an important caveat here, because of the inherent technical challenges associated with implementing noise infusion-based SDL solutions within a differentially private risk assessment framework, especially for complex weighted survey designs. Similarly, any particular SDL implementation (including those within a differential privacy risk assessment framework) may be better or worse at maximizing utility, depending on that implementation's design and the parameter choices reflected therein. Thus, the adoption of differential privacy as a risk assessment framework should be seen as entirely distinct from the selection of any specific SDL implementation (good or bad) informed by that risk assessment.}  Other \change{risk assessment} approaches also offer value, but need more research to address important shortcomings. 

\change{In a recent paper,   Hotz et al. \cite{hotzetal} called for a moratorium on the use of differential privacy. They argued that it is incompatible with  cost-benefit analyses of releasing new data, and that it prevents the computation of ``an essentially unconstrained
number of statistics'' \cite{hotzetal}.  The first claim is incorrect---it is contradicted by real-world deployments of differential privacy (including the 2020 Census---see Section \ref{sec:utility} and Appendix \ref{sec:si-utility}). As we explain in this paper, the second claim could be levied against \emph{any} method that controls disclosure risk (see, for example, criticisms of cell suppression in earlier censuses \cite{mckenna:2018}). Consequently, the question of whether ``an essentially unconstrained number of statistics'' should be supported is not a matter of choosing between differential privacy or an alternative disclosure avoidance technology. Instead it is a question of whether to protect confidentiality at all.}

\change{The quantity, granularity, and precision of published statistics all contribute toward the overall disclosure risk of the data. To protect confidentiality for its 100+ recurring censuses and surveys, the Census Bureau makes use of a wide variety of disclosure avoidance methods, including various implementations of suppression, coarsening, and noise injection approaches, including but not limited to those based on differential privacy. Because the application of any disclosure avoidance method reduces the utility of the resulting data in non-trivial ways, and because the choice of any particular disclosure avoidance method and how it is implemented can have differing impacts on utility for diverse subsets of use cases, the Census Bureau takes great care to select the best available disclosure avoidance method for each particular census and survey it conducts and for each data product it releases. These choices,   informed by external stakeholder feedback,   help maximize the resulting utility of published data for those uses that provide the greatest societal value. However, they are made with the understanding that protecting confidentiality requires that utility for some other potential or unanticipated use cases may suffer as a result. While these trade-offs may not be apparent for smaller data products (those with fewer statistics or less granularity), or for those that are used for only a narrow range of use cases, they will be visible and inescapable for large-scale, highly-granular data products with many diverse use cases, such as the 2020 Census.}



This paper is organized as follows. In Section \ref{sec:utility}, we \change{briefly review} activities performed by the Census Bureau to quantify \change{the} utility of 2020 Census data products.
In Section \ref{sec:desiderata}, we list some generally accepted requirements that disclosure risk measures should satisfy. Although they are simple enough to be discussed in a non-mathematical way, they are still rich enough to provide meaningful insights. In Section \ref{sec:methods}, we  overview \change{three} of the major methodologies for measuring disclosure risk. In Section \ref{sec:analysis}, we use the desiderata  to investigate the suitability of these disclosure risk assessment methodologies.  
We discuss implications for the federal statistical system in Section \ref{sec:discussion} and present conclusions in Section \ref{sec:conc}.
\change{The Appendices include: \ref{sec:si-utility}. Summary of how the Census Bureau used stakeholder feedback to assess how the SDL applied to the 2020 Census affected accuracy; \ref{sec:si-flexibility}. Comparison of different SDL methodologies that explains why the combination of aggregation and noise infusion was used for the 2020 census; \ref{sec:si-synthetic}. Comparison of synthetic data methods with traditional SDL; \ref{sec:si-costbenefit}. Discussion of cost-benefit and social welfare analyses of disclosure risk; and \ref{sec:si-errors}. Detailed analysis of published claims about the 2020 Census Disclosure Avoidance System and risk analysis.}

\section{Quantifying Utility via Stakeholder Engagement}\label{sec:utility}
Population censuses have many uses. It is unrealistic to quantify their social benefit with a single number. For the 2020 Census, the Census Bureau developed a suite of metrics, based on stakeholder engagement, that were used to understand the impact of different levels of accuracy on stakeholder use-cases. These metrics guided the development and tuning of the 2020 Census Disclosure Avoidance System (DAS) and clearly document the extensive balancing of interests that occurred \cite{tdahdsr}. See the details in Appendix \ref{sec:si-utility}. 

\section{Desiderata for  Disclosure Risk Assessments}\label{sec:desiderata}
Suspected privacy breaches for Census Bureau data products, protected using traditional methods, are routinely reported to the Chief Privacy Officer or the Disclosure Review Board. They are investigated but, as a matter of policy, the agency generally does not publicly confirm that a specific suspected breach is an actual breach. The agency has released some breach summaries \cite{mckenna:2019:reid}. 

Many breaches also go unreported---what is the incentive for an attacker to reveal a breach that is being exploited? Furthermore, reporting confidentiality breaches is sometimes discouraged by \change{curating organizations, e.g.,} IPUMS International \changetwo{(``[a]lleging that a person or household has been identified is also prohibited'' \cite{ipumsi})}. 

A typical argument used to justify weaker privacy protections is ``there have been no breaches so far.''  Such statements are often demonstrably false \cite{mckenna:2019:reid} and also illustrate the belief that loss of trust due to privacy breaches is easy to overcome and should not affect data quality. There is no evidence to support this belief, and strong evidence against it \cite{childs:et:al:2019}.

As one prominent state demographer recently pointed out at a Federal State Cooperative for Population Estimates meeting ``in past censuses, it was possible to create a PUMS-like \changetwo{[Public-Use Microdata Sample]} file from the SF1 tables \changetwo{[i.e., decennial census publications]}. It was just a pain in the neck'' (see Appendix \ref{sec:si-reconstruction}). Whether the reconstruction of microdata records for possible demographic or redistricting uses should be considered a privacy breach is a question requiring a nuanced response. However, the knowledge that it is attempted even by those who accept the requirements of confidentiality protection illustrates that the strengths and limitations of different methods of protection must be taken seriously, and that defensible methodologies for disclosure risk assessment are essential.

We now discuss desiderata that should be considered when assessing methodologies that quantify disclosure risk. It is important to note that a disclosure risk assessment methodology is not the same thing as a disclosure limitation mechanism. A mechanism  $\mech$ is a piece of software whose input is confidential data $\data$ and whose output consists of data products $\public$ that are suitable for public release. The role of the assessment methodology is to measure the  disclosure risk of the data products in a principled manner.  Disclosure risk assessment methodologies are not mutually exclusive, meaning that privacy properties of $\mech$ can be analyzed by multiple methodologies, though of course only methodologies that are defensible should be used for assessment.
The aim here is to present reasonably \change{un}controversial desiderata for risk assessment methodologies. Even with these simple desiderata, this exercise provides insights about various disclosure risk assessment methodologies. 


\subsection{Transparency}\label{sec:des:trans}
The principle of transparency is that details (e.g., source code and algorithm parameters) of a mechanism $\mech$ should be available to the public. The methodology for disclosure risk assessment should consider the possibility that an attacker would use the details of $\mech$ to attempt to reverse-engineer \cite{downcoding}  the public data products $\public$ to extract information about the confidential data $\data$ as, for example, can be done with many mechanisms supporting $k$-anonymity \cite{kanonymity}.

Transparency is necessary so that users of the public data products $\public$ can understand what errors\change{, limitations, or systematic} biases the data might have \cite[B1]{asa:2022}. This understanding facilitates public privacy vs. utility discussions and makes it possible to quantify utility through stakeholder engagement, as explained in Appendix \ref{sec:si-utility}.
Transparency is also necessary for the public to verify that $\mech$ was implemented correctly. A 2010 analysis by Davern et al. \cite{davern} revealed mistakes in the implementation of earlier SDL systems, with important implications for data quality. These mistakes were only detected because there were multiple sources of information available about the same population and estimates from the different sources were inconsistent. When multiple sources are unavailable, mistakes may not be detectable unless the code of $\mech$ is public. Abowd and Schmutte \cite{abowd_schumtte:2015} call this property of SDL ``discoverable''  and point out that unless the system is either discoverable or public there is a strong risk that the likelihood function of the released data \change{(the probability distribution used by statisticians to make inferences)} suffers from nonignorability (\change{a systematic bias like} nonignorable missing data \cite{little:rubin:2002}), making correct scientific inferences difficult.

Many uses of SDL are \emph{not} transparent. In particular, the 2010 Census used non-transparent methods (i.e., data swapping \change{with a confidential algorithm and parameters}) due to concerns that knowledge of implementation details could be used to undo the privacy protections. This resulted in a general lack of awareness that data had been modified. There is little understanding of how non-transparent SDL affects analyses (in particular, small area estimation \cite[e.g., Table 1]{abowd:hawes:2023}). In contrast, in the 2020 Census, the Census Bureau embraced transparency, publishing source code of the 2020 DAS 
\cite{prodredgithub}, providing demonstration data products using historical data \cite{ddp}, developing and releasing associated summary metrics, and triggering a robust debate on the proper role, and balance, of privacy and utility in the federal statistical system \cite{hdsrissue}. 

\subsection{Universality}\label{sec:des:universal}
A method for disclosure risk assessment should be applicable to any type of data product: microdata samples, tabulations, regression coefficients, etc., even if the agency only plans to release a single product because all data products contribute to disclosure risk, and their cumulative impact must be assessed when, inevitably, additional products are proposed from the same source. While some proposals call for the release of only microdata, restricting an assessment method to only work on microdata is too limiting. For example, researchers may wish to publish regression coefficients from a Federal Statistical Research Data Center (FSRDC). In general, the format of a data product should be decided as part of the privacy/utility cost-benefit analysis, and a disclosure risk assessment methodology should support this analysis by providing a privacy cost for any data product format. 


\subsection{Deem releasing uninformative statistics not a disclosure risk}\label{sec:des:meta}
Consider a data release that produces something completely \change{un}informative, like a heavily perturbed total population count. This information should be considered a minimal disclosure risk. This is a sanity check but, as we shall see, some disclosure risk assessment methodologies fail it. 



\subsection{Support \underline{arbitrary} \change{statistical} analyses}\label{sec:des:arbitrary}
The types of questions that researchers and policymakers \change{try} to answer often involve complex data analyses. This raises an important concern---would a disclosure risk assessment methodology stand in the way of these activities? We split this issue into two desiderata: (1) a disclosure assessment method should allow the creation of data products that accurately support an arbitrary, \emph{pre-specified} \change{statistical} analysis, and (2) the assessment method should allow the creation of data products that support arbitrary analyses that have \emph{not} been specified in advance. We discuss the first desideratum here and the second desideratum, which turns out to be equivalent to supporting arbitrarily many \change{kinds of statistical} analyses, \change{in \ref{sec:des:many}}.


Some analyses are inherently incompatible with any reasonable conception of privacy---for example, analyses designed to determine the identities of a specific person's sexual partners or parties to payments for illicit goods. A public health authority or law enforcement agency, respectively, might have legitimate reasons to undertake such investigations, but such activities are clearly prohibited for a U.S. statistical agency by the definition in Statistical Policy Directive 1 (SPD 1, 44 U.S.C. § 3561(8)) of ``nonstatistical purpose.''
Any pre-specified \emph{statistical} analysis--for example, aggregate counts, regression designs, learning models---for a \emph{large enough} population should be supportable.

This desideratum includes tiered access and restricted \change{ data use} environments like the FSRDCs, meaning an analysis need not be pre-specified at the time of the creation of a data product.
%
%
%
Instead, researchers can analyze public data products and when preliminary results indicate potentially useful findings (or that a potentially useful investigation cannot be supported by existing data products), the researchers can specify the analyses they need to perform and  use the FSRDCs to conduct them. The results of the work in the FSRDC can then be protected with privacy technology and publicly released\change{,} as Chetty et al. did with IRS and Census Bureau data \cite{chetty:et:al:2018} and with Facebook data \cite{chetty:et:al:2022}.
\changetwo{Note, however, that quantifying privacy loss due to exploratory analysis of raw data, known as the ``Fienberg Problem'' \cite{fienbergproblem}, is an open research question.} 


\subsection{Support arbitrarily \underline{many}  analyses}\label{sec:des:many}
Given the explosive growth of various research communities with different needs, goals, and statistical methodologies, stakeholders may wish to avoid FSRDCs and would be delighted to receive data products that support arbitrary  analyses that are not specified in advance. Certainly, this would be less cumbersome for them.
Since the analyses are not known in advance, the only way to support them at the time of data product creation is to ensure the data products can support arbitrarily \emph{many}  analyses.
However, providing a data product that accurately supports arbitrarily many analyses is mathematically equivalent to releasing confidential data $\data$ with no protections at all \cite{Dinur2003}---in other words, the confidential data could be reconstructed from those data products. 
%
Therefore neither differential privacy nor any other methodology for effectively assessing disclosure risk can legitimately satisfy the criterion of accurately supporting arbitrarily \emph{many}  analyses for any non-trivial degree of confidentiality protection \change{-- it is a mathematical impossibility.
}



\subsection{Deem nearly perfect reconstruction of high-dimensional data or nearly perfect inference about a data subject high disclosure risk}\label{sec:des:recon}

This desideratum acknowledges that data products that are nearly equivalent to the original confidential microdata (i.e., if microdata can be accurately reconstructed from the released product) is a disclosure risk if the original microdata were considered confidential. Note that, as discussed earlier, this principle is in conflict with support for arbitrarily many  analyses.

This is true even for data collected from sample surveys. In addition to enabling inference based on small homogeneous groups, near-perfect reconstruction renders sample uniques easily identified, and sample uniqueness is often a strong predictor of population uniqueness. The challenge of detecting and protecting a population unique---a record in the response data that represents the only entity in the population with the associated subset of characteristics---has a long history of study in surveys and censuses \cite{bethlehem:et:al:1990} as well as in biostatistics \cite{homeretal}. \change{Prior} work shows particularly strong results without requiring access to the confidential data for verification \cite{rocher:et:al:2019,dworkarsa}.

Reconstruction of an entire dataset is not necessary for a prohibited disclosure to occur. For example, suppose a dataset reveals that all 10 people in zip code 12345 who are 43-year-old Asian Hispanic males have cancer. The confidentiality of these individuals has been breached even if  the enumeration was \changetwo{im}possible. 

\subsection{Deem generalizable inference \change{not a violation of} confidentiality}\label{sec:des:gen}
 Dwork and Pottenger \cite{dworkpottenger} proposed explicitly acknowledging that generalizable statistical inference is not a privacy breach because it does not jeopardize the confidentiality of the underlying data. Data products are published to support generalizable inferences. Statistical inferences from the data can still be harmful, but the harm from such inferences should not be assigned to the SDL system. In the case of sample surveys, for example, one hopes that the findings from those data generalize to the entire population. In the case of censuses, one hopes that the results generalize to subsequent years.  For example, a policy implemented on the basis of the 2020 Census should still succeed in achieving its primary aims as the population grows and changes. 

Thus, one should expect to learn something new from the data, and this knowledge should not be considered a  breach of confidentiality. A canonical example \cite{dworkpottenger} is the large scale cancer study CPS-I spanning the 50s, 60s, and 70s \cite{cps1,hammond:horn:1954}, which conclusively established the link between smoking and cancer.
This finding means that insurance rates for smokers may have increased after the study was published, which would be an undesirable outcome from their point of view. However, if a smoker was not even a study participant, should this improved inference about their cancer risk be considered a privacy breach? According to this desideratum, it would not, since this inference was clearly based purely on generalizable knowledge. 

Now suppose a different smoker did participate in the study. Inferences about that person's cancer risk would result from a combination of generalizable knowledge obtained from the data and use of the subject's data in the production of the report. By accepting this desideratum, one is expecting a disclosure risk assessment methodology to be able to determine how much of an inference is due to generalizable knowledge and how much is due to participation in the dataset on which the inference is based.

There is a separate policy concern regarding the publication of data on sensitive subjects or populations. For example, if the Census Bureau learned that the main users of one of its data products were foreign intelligence agencies trying to manipulate U.S. elections, it would probably stop releasing this data product on policy grounds, even if the foreign intelligence agencies were only doing generalizable statistical analysis and no prohibited disclosure occurred. That is correctly a policy decision independent of the concerns about confidentiality but consistent with the guidelines in SPD 1 (44 U.S.C. § 3561-4). 

Rejecting this desideratum necessarily conflates decisions on appropriate statistical uses with those concerning disclosure limitation and risks harming the scientific value of public data sets.

\subsection{Support composition}\label{sec:des:composition}
Imagine a scenario in which one set of data products was deemed a low privacy risk, another set of data products was also deemed low privacy risk, but their joint release was deemed to be a catastrophic privacy violation.
The technical terminology is that the two privacy-protected data products did not \emph{compose} well--the cumulative effect of all publications from the same confidential data source(s) contributed to the disclosure risk in ways that could not have been predicted from separate, independent analyses \cite{ganta2008composition}. Separate ``release and forget'' analyses have been standard practice for SDL practitioners in the past, as was the case for disclosure risk reviews done on data products from the 2010 Census \cite{mckenna:2018,mckenna:2019:microdata}, but they provide no indication of the cumulative disclosure risk of all publications, unless the disclosure assessment methodology supports composition.

A disclosure risk assessment method composes well when an upper bound on the joint privacy risk for a set of data products is predictable based on a privacy analysis conducted on each data product separately. This property simplifies the design of disclosure limitation mechanisms and allows different parts of an organization to plan different product releases with minimal coordination as long as there is centralized review and control of the cumulative disclosure risk.
It also simplifies the marginal cost-benefit analysis of each new data publication. For instance, one could separately analyze the incremental privacy cost of each new data publication and weigh it against its incremental utility. 

Rejecting this desideratum means organizations must plan their data releases sequentially, measuring the cumulative effect of all prior releases from the same confidential source, as well as the proposed new data release. This will affect the timeliness of data product releases because an organization would release one data product, then must conduct an analysis of what happens when a second data product is added. Subsequently, \change{when planning} a third data product, it must be jointly analyzed with the first two. Adding each new data product becomes an increasingly complex endeavor. This is a limitation that, for example, currently impacts all statistical products that use properly implemented primary/complementary suppression, as discussed in \cite{abowd:hawes:2023}.

\subsection{Acknowledge multiple types of attackers \change{and auxiliary information}}\label{sec:des:multiple}
When publicly releasing data products, privacy risks can come from a variety of sources. For example, \change{information technology} giants may combine the released data with their own user data to obtain information they did not previously possess, as their data can differ in quality, extent, and measured attributes when compared to carefully worded survey data. Another source of privacy risk comes from data brokers who buy and sell data about individuals in bulk, companies that \changetwo{engage} other companies to access their customers and data, prying neighbors, redistricters who want  personal data as they can to predict individual voting patterns, or government agencies that are enjoined by law from requesting or issuing subpoenas for the confidential data.

The acknowledgment and analysis of privacy risks due to multiple different attackers (who do not necessarily share data) is critical to avoid mistakes. For instance, if the only type of attacker considered is an equivalent of Amazon 
(e.g., has names,  addresses, and purchasing histories of a large portion of the population), then a disclosure limitation methodology based solely on that attacker could lead to the erroneous conclusion that publishing addresses and consumption patterns of survey participants does not introduce any risks since the data are ``already out there.'' \change{Other attackers may not have Amazon's knowledge, and SDL should prevent them from learning personal information they don't already have even though it is ``out there.''} 

\subsection{Resist brittleness}\label{sec:des:brittle}
\change{A disclosure risk assessment exhibits brittleness if the inclusion or removal of data about one entity materially compromises the privacy of another entity's data. Confidentiality protection for businesses can be brittle in this sense because whether or not a competitor supplied data can affect the confidentiality protection afforded to other businesses in the same market. An agency does not exclusively control} information about any particular data subject---the data subject can do whatever they want with their \change{own} information. Thus a disclosure risk assessment method should anticipate such a possibility and should not be brittle---whether or not a data subject decides to release their own data \change{publicly or to the agency} should not materially affect the assessed privacy risks of everyone else.

\subsection{Be resilient to misspecification}\label{sec:des:mis}
Many disclosure risk assessment methodologies make assumptions about capabilities of the attackers---the data they may or may not possess, the probability distribution that encapsulates their prior beliefs, the attack method they may choose to use, and even which pieces of information should be considered sensitive.
What if the assumptions are wrong \changetwo{ or obsolescent (e.g., when new external datasets become available in the future)}? This is similar  to model misspecification. Disclosure risk assessment methods should be able to quantify how they are affected by deviations from their underlying assumptions.

\subsection{Control computational effort}\label{sec:des:comp}
As we saw with composition, computational effort can become a major practical consideration when creating data products. Computational expenses can be incurred when \change{including} all prior data products in a disclosure risk analysis, when computing posterior probabilities in Bayesian methods, and generally when handling large data sets. Ideally, reducing computational expense should not come at the cost of weakening the disclosure risk assessment (e.g., making assumptions about prior distributions that are clearly incorrect, but allow for closed-form estimates).

\subsection{Control manual effort}\label{sec:des:manual}
The development of SDL systems typically requires expertise in statistical modeling, computational algorithms, and optimization. The use of a disclosure risk assessment methodology incurs an initial cost in staff training\change{. It} may also incur marginal costs in keeping track of data that are available elsewhere (and that may be used by an attacker) and analyzing disclosure risks of new data products. It should also not hinge on the assumption that personnel are more adept at finding viable attack vectors than actual potential attackers.
Keeping the manual effort manageable is another important  goal for any statistical agency.

\section{Major Disclosure Risk Assessment Methodologies}\label{sec:methods}

Having identified \change{several} major desiderata for evaluating prospective disclosure risk assessment methodologies, \change{we next} consider some of the major methodologies that have been proposed for this risk assessment: absolute disclosure risk, prior-to-posterior comparisons (a type of \emph{relative} disclosure risk), and counterfactual posterior-to-posterior comparisons (another type of \emph{relative} disclosure risk). To keep the mathematical detail at a minimum, only the motivations and broad outlines are described, but this is still enough to compare the different frameworks along the desiderata outlined in Section \ref{sec:desiderata}.

\subsection{Absolute disclosure risk}

Absolute disclosure risk can be formulated in different ways. \change{For concreteness, we follow a} specific proposal from Hotz et al. \cite{hotzetal} \change{in} the following. The confidential data set $\data$ consists of records of $n$ data subjects $\person_1,\dots,\person_n$. Each person $\person_j$ has sensitive information $\sensitive_j$. An attacker has some data $\attackinfo$ (e.g., an external dataset such as voter registration records, or personal observation) \change{ and a Bayesian prior, $P(s_j = S \text{ and record } r_i \text{ belongs to person } p_j)$}. Since this risk assessment approach requires linkage to external information, it also requires that the data product $\public$ produced by mechanism $\mech$ be either a dataset of records $\{\precord_1,\dots, \precord_n\}$ (e.g., a public microdata file), or that a partial or complete microdata file could be reconstructed from non-microdata data products (e.g., reconstruction of a dataset of records $\{\precord_1,\dots, \precord_n\}$ from tabular data releases). 

To attack a person $\person_j$, the attacker must first determine which protected record $\precord_i$ corresponds to that person and then must infer a value $\svalue$ for $\sensitive_j$, the sensitive information about $\person_j$.
For discrete attributes, the probability of success is written as 
$P(\sensitive_j=\svalue \text{ and }\precord_i \text{ belongs to person }\person_j ~|~\attackinfo, \public, \mech),$
where conditioning on $\mech$ indicates that the mechanism is known but the realized values of its random variables, if any, are not known.
Computing this probability requires making assumptions about the attacker's prior knowledge $\attackinfo$ and prior beliefs about which distribution(s) could have generated the data. We refer to this probability as the \textbf{absolute disclosure risk with linking}.

Hotz et al. \cite{hotzetal} \change{present an analogy to pathology}, in which this absolute disclosure risk is analogous to the probability that someone will die within $x$ months---clearly a number that people would care about.
We  extend this analogy when discussing prior-to-posterior and counterfactual comparisons as it presents a vivid way of highlighting the differences in methodologies.

We also consider a modification of this proposal that we refer to as \textbf{absolute disclosure risk without linking}.  In this setting, an attacker only needs to learn the sensitive information $\sensitive_j$ about person $\person_j$ without having to specifically identify which privacy protected record $\precord_i$ belongs to $\person_j$. As demonstrated by the attack of Homer et al. \cite{homeretal} on genomics data, linking is not necessary for serious disclosures to occur because the probability of correct attribute inference can be accurately estimated from other properties of the data \cite{dwork:et:al:2015}. Dropping the linking requirement also adds more flexibility since a mechanism $\mech$ no longer needs to produce a \change{record-level} dataset as the public data product $\public$---it can produce regression coefficients, tabulations, a deep learning model, etc. The success probability of the attacker can then be written as
$P(\sensitive_j=\svalue \text{ and person  }\person_j \text{ is in }\data ~|~\attackinfo, \public, \mech).$

\subsection{Prior-to-posterior comparisons}
Continuing the analogy to pathology, while absolute disclosure risk corresponds to the overall probability of dying within $x$ months, prior-to-posterior comparisons are comparisons between the current probability of dying and the probability of dying after a treatment or other activity, thus trying to isolate the effect of the treatment. Here, the treatment is the use of $\mech$ to release privacy-protected data products. 

In SDL applications, the attacker starts with a prior probability that the sensitive attribute $\sensitive_j$ of person $\person_j$ equals some value $\svalue$. Mathematically, the prior is
 $P(\sensitive_j=\svalue \text{ and person }\person_j \text{ is in }\data ~|~ \attackinfo)$.
After the confidential data $\data$ are run through $\mech$ to produce the public data products, the attacker can form the posterior distribution:
$P(\sensitive_j=\svalue  \text{ and person }\person_j \text{ is in }\data~|~ \attackinfo, \public, \mech)$.
These two probabilities can be compared, typically, but not always, by subtraction: $\text{posterior}-\text{prior}$, or division: $\frac{\text{posterior}}{\text{prior}}$ to measure the change in beliefs induced by the public data products. Because such disclosure risk methods are comparisons (i.e., before and after $\public$ is released), they are known as measures of \emph{relative disclosure risk}.

\subsection{Counterfactual posterior-to-posterior comparisons} This framework \changetwo{is} used by differential privacy and its variants \cite{BassilyGKS13}. Consider a clinical trial \changetwo{where} the treatment group receives  medicine for six months, and a control group receives a placebo for six months. Comparing the mortality of these two groups is analogous to counterfactual posterior-to-posterior comparisons \cite{rubin:1974}.
In terms of disclosure limitation, this methodology considers two ``worlds''---the \emph{actual} world in which the target person's data are included in the dataset $\data$ that is sent to $\mech$, and the \emph{counterfactual} world in which that person's data are omitted or replaced by a blank record before being sent to $\mech$.
The idea behind the counterfactual world is that a person may have a set of actions that can be performed unilaterally to protect privacy, such as  not participating in a survey or submitting a blank response. If an individual considers these actions sufficient to protect their privacy, then the counterfactual world is a privacy-preserving baseline for that person.

The actual world and counterfactual world can be compared in several ways. The first, often called the \emph{frequentist approach}, is developed from the perspective of hypothesis testing: based on the data products $\public$ and the source code of $\mech$, could the attacker detect if the full dataset was the input to $\mech$ or whether the target person's record was removed before running $\mech$ (i.e., could the attacker distinguish between the actual and counterfactual worlds)?

The second way of comparing the actual and counterfactual worlds is often called the \emph{Bayesian approach}. One would compare the inference that an attacker would have in the actual world:
$P(\sensitive_j=\svalue ~|~\attackinfo, \public, \mech,$ \text{ target} $\person_j$ is in the data) 
to the inference the attacker would have made in the counterfactual world:
$P(\sensitive_j =\svalue ~|~\attackinfo, \public, \mech,$, $ \text{ target $\person_j$ was \emph{removed} from the data})$.
If the two conditional probabilities are similar it means that the attacker \change{is likely to} draw nearly the same conclusion even if the target individual had never participated in the survey. The two conditional probabilities can be compared by dividing one by the other, but more sophisticated comparisons are often used instead.
A detailed  treatment  can be found in \cite{semanticspaper}.

\subsection{Discussion}
Recalling that a disclosure risk assessment methodology is not the same thing as a mechanism $\mech$, consider a mechanism that adds noise to the sufficient statistics of a model and samples records from the model to produce a synthetic dataset. The overall effectiveness of this mechanism at mitigating disclosure risk can be analyzed from the perspective of all three disclosure risk assessment methodologies discussed above. 
%
Thus, common SDL methods like suppression, coarsening, swapping, etc. \cite{sdcpractice}, 
are not discussed here as risk assessment methodologies since they are frameworks for designing mechanisms which can then be analyzed by disclosure risk assessment methodologies.
We also note that each methodology is an umbrella that may have many variations depending on the definition of sensitive information, the types of probability distributions used by the attacker, the knowledge the attacker can possess, and how probabilities can be compared to each other.

It is often claimed  that privacy-loss measures obtained from the counterfactual methodology place a hard limit on the privacy loss---that is, after a certain point, there can be no further interaction with the underlying confidential data. On the contrary, the counterfactual method and \emph{all} other methodologies provide a measure of disclosure risk. What an agency does with the risk measures is up to the agency. It can weigh the incremental privacy risk against the incremental gain in utility of subsequent data publications. Any claims that rigid, dogmatic restrictions are pre-requisites to the use of various disclosure risk assessment methodologies are specious---the only requirement is that an agency must make decisions that are defensible.

\section{Analysis of Methodologies}\label{sec:analysis}

We next compare the three different methodologies from Section \ref{sec:methods} using the desiderata from Section \ref{sec:desiderata}. We go through the desiderata in the same order. See Table \ref{tab:summary} in Section \ref{sec:analysis:summary} for a summary. The purpose of these desiderata is to help ensure that a methodology is chosen based on sound principles, but it also helps identify promising areas of future privacy research.

\subsection{Transparency}
In principle, all three methodologies \emph{could} support transparency, and so the main question is whether the methodologies have a viable proof of concept. There are two main challenges for a proof of concept. \emph{Computational}---can the relevant posterior probabilities $P(\cdot ~|~\public,\mech, \attackinfo)$ be computed accurately and efficiently enough for decision-making purposes, even when $\mech$ is complex? \emph{Model specification}---where do the attacker's probability distributions and information come from, since computing the posterior distribution requires specifying $P(\data)$, the prior probability of the data. If the data curator must estimate $P(\data)$ from the actual data, then the methodology could become non-transparent because the data curator would either have to reveal $P(\data)$, so that the public could verify the disclosure risk assessment, or would have to hide it, resulting in degraded transparency. If the data curator reveals a $P(\data)$ estimated from the data, then this could itself lead to disclosure.

The methodology based on counterfactual posterior-to-posterior comparisons has many proofs of concept, most notably the literature on differential privacy. A comprehensive survey on the different ways of measuring disclosure risk under this methodology can be found in \cite{sokdps}. There are many computationally-friendly tools for accurately approximating or upper-bounding the privacy loss in this framework \cite{zcdp,
pmlr-v151-zhu22c}. Interestingly, even though posterior distributions may be difficult to compute by themselves, one only needs to \emph{compare} two posterior distributions (for the actual and counterfactual worlds) and this comparison can be performed efficiently. The other challenge, model specification, is typically resolved by providing protections for absolutely all priors. For decision-makers who are used to analyzing re-identification probabilities, the counterfactual contrast can be converted into significance level and statistical power for the null hypothesis of determining participation in the dataset (a pre-requisite to re-identification) versus the alternative of non-participation  \cite{wasserman:zhou:2010,semanticspaper}. 
%
Such an approach uses trade-offs that statistical agencies have long employed in designing surveys. 
Other approaches use a smaller set of priors \cite{Hall_Wasserman_Rinaldo_2013,sokdps}, but their strengths and weaknesses require further study. This is one promising direction of research for this methodology. Thus the counterfactual framework  supports transparency through many proofs-of-concept.

The prior-to-posterior comparison methodology also has some transparent proofs-of-concept \cite{pufferfish,wassermech} but they are more challenging to implement. Model specification is handled by providing guarantees for a large set of priors or for priors that are easy to work with (such as the Gaussian distribution). Aside from the ``easy'' priors, which may not faithfully represent the data, these methods also encounter computational obstacles when computing posterior distributions. Monte-Carlo and MCMC simulations can often be used for these purposes \cite{mcclure}. However, for large-scale data sets, MCMC approaches can be slow to converge, potentially leading to inaccurate estimates of disclosure risk.

The absolute disclosure risk methodology, however, currently lacks a proof-of-concept. When the public data products $\public$ take the form of microdata, absolute disclosure risk with linking is typically measured by performing linking attacks between the confidential data $\data$ and $\public$ \cite{reiter:2005}.
Such empirical assessments, however, are frequently criticized for using simplistic, rather than the best available, linking algorithms \change{ and external data}, and often significantly \emph{underestimate} disclosure risk  \cite{rocher:et:al:2019}. In addition, they reduce transparency since the results, upon which decisions are based, cannot be audited by the public because such audits require access to the confidential data, which could, however, be granted to an auditor within a restricted-access environment. Absolute disclosure risk \emph{without} linking faces the two open challenges: specifying a prior distribution $P(\data)$ that would not lead to underestimation of disclosure risk, and computing the posterior distribution efficiently and accurately in large-scale data.
\change{Thus, currently only the prior-to-posterior and counterfactual methodologies have transparent proofs-of-concept, the latter being more numerous.}

\subsection{Universality}\label{sec:analysis:universal}
The absolute disclosure risk without linking, prior-to-posterior, and counterfactual methodologies all place no restrictions on the public data products $\public$. Only absolute disclosure risk \emph{with} linking restricts the form of $\public$, because it requires the released product to be microdata or of a form that could be reconstructed into microdata. Thus, it would not be applicable to many common alternate forms of data releases, such as regression coefficients, that might be released from an FSRDC.

\subsection{Deem releasing uninformative statistics not a disclosure risk}\label{sec:analysis:meta}
Consider a data collection effort that records personal information about individuals such as race, citizenship, etc. Releasing a ``data product'' that consists \change{\emph{only}} of a heavily perturbed national population count would be treated as nearly zero disclosure risk under the prior-to-posterior methodology---the posterior probability of learning sensitive information after releasing such data is nearly the same as the prior probability of the sensitive information due to the non-informative data product. Similarly, it has nearly zero disclosure risk under the counterfactual framework (see the explanation of per-attribute semantics in \cite{semanticspaper}).

Absolute disclosure risk with linking is not applicable in this setting, but absolute disclosure risk \emph{without} linking could be applied. In some situations, the absolute disclosure risk method would assign a large disclosure risk to the national count. For example, a target individual's neighbor may be completely certain that the target participated in the census and may have a very good guess about the target's race, marital status (e.g., same-sex or opposite-sex), etc. In some cases, even citizenship status would be entirely predictable for certain people.
In other words, the prior belief
$P(\sensitive_j=\svalue \text{ and person }\person_j \text{ is in }\data ~|~ \attackinfo)$ is already large.
The release of uninformative statistics would not change the beliefs and so the posterior  $P(\sensitive_j=\svalue \text{ and person  }\person_j \text{ is in }\data ~|~\attackinfo, \public, \mech)$, which is the only quantity this methodology considers, would also be large. Therefore, the heavily perturbed count would not be releasable \emph{even though the neighbor learned virtually nothing new about the target individual}. See also the Montana example in Section \ref{sec:analysis:gen}. 

\change{This desideratum suggests that a posterior probability needs to be compared to \emph{something} in order to determine whether a data release changed an attacker's beliefs or reduced their uncertainty. Absolute disclosure risk specifically avoids such comparisons.}

\subsection{Support \underline{arbitrary} \change{statistical} analyses}
\label{sec:analysis:arbitrary}
Tiered access can minimize the need for pre-specifying all potential uses of data in advance. Analyses that are formulated later, and which are not supported by public data products $\public$,
can then be performed in restricted access services like FSRDCs with the outcomes subsequently protected by SDL.  While this can affect the pace of research, it currently appears to be unavoidable.
In principle, absolute disclosure risk without linking, prior-to-posterior comparisons, and counterfactual comparisons can be used to analyze the privacy risks of the outcomes produced in restricted-access services, but absolute disclosure risk with linking may not be applicable.

When the discussion tilts towards known mechanisms $\mech$ for different kinds of analyses, and whose privacy risks can be efficiently assessed, the counterfactual methodology has the most support \change{and underlying research at the moment.}
\change{Nevertheless, supporting a wider variety of analyses is an important research direction.} 

\subsection{Support arbitrarily \underline{many}  analyses}
\label{sec:analysis:many}
The difference with the previous desideratum is the quantity \changetwo{and} variety of arbitrary analyses. As mentioned in Section \ref{sec:desiderata}\ref{sec:analysis:many}, the only way to allow  researchers to conduct arbitrarily many  analyses accurately on a public data product $\public$ is to violate confidentiality and simply make $\public$ identical to $\data$. This mathematical fact \cite{Dinur2003} must be acknowledged. No disclosure control methodology can achieve this goal.

\subsection{Deem nearly perfect reconstruction of high-dimensional data or nearly perfect inference about an individual high disclosure risk}\label{sec:analysis:recon}
When the confidential data $\data$ can be nearly perfectly reconstructed from the public data $\public$, absolute disclosure risk without linking, prior-to-posterior, and counterfactual comparisons would assess $\public$ as a high disclosure risk. When $\public$ takes the form of microdata, then in certain situations, absolute disclosure risk \emph{with} linking will not characterize this release as high risk. For example, suppose the target individual is known to be in the dataset (or the dataset may be a full census). If the target individual shares the same linking attributes with 9 other people, and if all of their sensitive information is the same, then an attacker has learned the target's sensitive data without knowing which of the 10 records belongs to the target.
This sensitive attribute inference is probably not a generalizable inference because it is based on a sub-population of only 10. It is the type of inference that can raise a concern about confidentiality, as noted for k-anonymity \cite{kanonymity}. In a sense, the inability to link seems irrelevant, but absolute disclosure risk with linking discounts the perfect inference specifically because of the limitations on linkage, ignoring the success of the non-generalizable inference. This is another example of a simple situation that is not handled correctly by absolute risk assessment with linking, casting doubt on its proper behavior in more complex situations.

\subsection{Deem generalizable inference \change{not a violation of} confidentiality}
\label{sec:analysis:gen}
In the counterfactual methodology, using $\public$ to make inferences about a target who is not in the data (the counterfactual world) is considered to be inference based on generalizable statistical knowledge. On the other hand, inference about a target who \emph{is} in the data (the actual world) depends on both generalizable knowledge and the use of their records in creating $\public$. Thus comparison of the two attempts to isolate the specific risk due to being in the data from learning something new about the population.
On the other hand, if the target person is in the data, absolute disclosure risk does not attempt to distinguish between whether an inference about sensitive information $\sensitive$ of the target is primarily due to generalizable statistical knowledge or the target person being in the data. This harms data utility since absolute disclosure risk methodologies would limit or suppress both types of inferences.

These are not just theoretical concerns. The inability of absolute disclosure risk to factor out generalizable inference can harm demographic analysis. For example, suppose that the self-reported race and ethnicity of individuals (or of minority populations)  needs to be protected. There are significant amounts of homogeneity in some block groups, tracts, counties, and even states---regions with sufficient populations for generalizable inference. For example, in 2010, 89\% of the population of Montana was ``White Alone,'' but there are clusters of minority populations in its counties, tracts, block groups, etc. Knowing that someone lives in Montana, or one of the homogeneous smaller regions, would allow race inference with at least 89\% accuracy---this is the kind of statistical inference that demographers want to be able to make. Under absolute disclosure risk assessments, this accuracy would be considered a threat and the demographics of these regions would need to be suppressed. However, anyone who is aware that suppression was done for homogeneous regions (e.g., regions with 89\% homogeneity) would know these regions are homogeneous and would likely be able to guess the prevalent race/ethnicity combinations. Thus, additional complementary suppression would be needed to add uncertainty about why a region's statistics were suppressed. On the other hand, under counterfactual reasoning or careful prior-to-posterior comparisons, this would not be a risk. In particular, the counterfactual methodology seeks to protect how an individual differs from their reference population. In other words, once you know that Montana is 89\% White Alone, the counterfactual approach would seek to protect inferences about how any individual differs from this baseline.

Prior-to-posterior comparisons can be set up to remove the component of inference that is due to generalizable statistical knowledge \cite{pufferfish}. One way to do that is to have the attacker use the true data-generating distribution in their attack. However, since  the true distribution is generally unknown even to the data custodian (who would need it to simulate the attack), an alternative is to use attackers who have complete confidence in their beliefs about data distributions (i.e., they do not put priors on hyperparameters \cite{bda}). In this way, the only difference between the prior and posterior distributions would be due to the use of the target person's record as part of the input to $\mech$.

\subsection{Support composition}\label{sec:analysis:composition}

Composition is well-studied in differential privacy \cite{diffpbook}. The counterfactual methodology strongly supports this property.
On the other hand, much less is known about composition in the prior-to-posterior framework \cite{wassermech} and no known instances of the absolute disclosure risk framework satisfy composition. This is an area where more research is needed. 
Lack of composition could lead to undesirable situations in which a planned data product must be scrapped because it interacts with previously released data products, even though each of the data products were assessed to have low disclosure risk on their own. Composition has significant practical importance to statistical agencies.

\subsection{Acknowledge multiple types of attackers \change{and auxiliary information}}\label{sec:analysis:multiple}
The counterfactual framework \cite{dpsemantics} and prior-to-posterior comparisons \cite{pufferfish} defend against multiple attackers. However, it is problematic for absolute disclosure risk. As discussed in Section \ref{sec:analysis}\ref{sec:analysis:meta}, nearly every individual can have high absolute disclosure risk when the attacker is their neighbor. The disclosure may not be due to the release of $\public$, but the absolute disclosure risk methodology does not compare any \emph{changes} in beliefs that result from $\public$. Thus, consideration of multiple attackers under the absolute disclosure risk framework could result in no data being released at all, or it could be compromised by the removal of many realistic attack scenarios. This again motivates using the other methodologies to make decisions about the privacy utility trade-offs when considering mechanisms for creating public data products.

\subsection{Resist brittleness}\label{sec:analysis:brittle}

For mechanisms like principled suppression \cite{sdcpractice}, if a few individuals reveal their data, it can dramatically reduce the privacy protections for other individuals. This is an example of brittleness. Counterfactual comparisons assign lower disclosure risk to mechanisms that are not brittle. In fact, one of the guarantees of differential privacy is that even if all other individuals reveal their data, the target individual is still protected \cite{diffpbook}. Not every type of prior-to-posterior and absolute disclosure risk approach resists brittleness. However they can be made to support this property \cite{mcclure} by adding information about other individuals into the attacker's information set $\attackinfo$. 

Brittleness also involves situations where some unperturbed statistics about the data have been released. One example is the release of the as-enumerated state population counts in the 2020 Census \cite{tdahdsr}. It is commonly believed that this irreparably breaks the differential privacy guarantees \cite{hotzetal}, but this is not the case---all it does is change the type of counterfactual analysis that is possible. To analyze the situation where the state population totals are invariant, one would use the following setup. In the counterfactual world, the target individual would submit their correct state, but would be free to alter everything else about their response. One would then compare the inference about the target in the world where their responses were truthful to the inference in this counterfactual world. Once the counterfactual world is properly chosen, the technical details of the analysis follow the examples in \cite{semanticspaper}. 


\subsection{Be resilient to misspecification}\label{sec:analysis:mis}

Disclosure risk assessment methodologies rely on attacker posterior distributions $P(\cdot~|~\attackinfo,\mech,\public)$, which require specifying the attacker's prior belief in the data-generating distribution: $P(\data)$. If one misspecifies this prior in a disclosure risk analysis, and if the true distribution differs from the specified distribution, how can one quantify the error in the disclosure risk assessment?
In the counterfactual framework, the most common variations of differential privacy are immune to this problem because they provide guarantees for all attacker priors \cite{dpsemantics,semanticspaper}.
In the prior-to-posterior framework, this issue is not well-studied, but Song et al. \cite{wassermech} showed that, in some cases, the effect of misspecifying the prior can be quantified using the Wasser\change{stein} distance between the true and specified distributions. Results for absolute disclosure risk are currently unknown.

\subsection{Control computational effort}\label{sec:analysis:comp}

\change{Reducing computational burden without decreasing the quality of the disclosure risk assessment should be an ongoing area of research.}
All the methodologies can be computationally \change{demanding}, especially for large-scale datasets. This is particularly true for methodologies that do not support composition (\change{e.g.,} absolute disclosure risk and most prior-to-posterior comparisons) since they must \change{include analysis of} external datasets and previously released data products \change{in order to avoid underestimating disclosure risk}.  \change{Such analyses may already be infeasible due to the vast quantities of available data.} 

\subsection{Control manual effort}\label{sec:analysis:manual}
All the methodologies require significant manual effort/technical skills, but, aside from statistical sophistication, where this effort is spent may differ. 
For the counterfactual framework, it is necessary to understand the complexities of differential privacy, and non-trivial effort is required in the design of mechanisms. However, there is also research into automating the design of mechanisms, known as \emph{program synthesis}, and ensuring that the disclosure risk assessment is mistake-free, known as \emph{verification} \cite{SA19}.
For variations of the prior-to-posterior comparisons that support composition, the manual effort is generally the same. For the other variations, the staff at a statistical agency would need to constantly find new sources of data that would be available to an attacker,  keep up-to-date with the latest types of attacks, and develop their own attacks as well. Similar requirements apply to the absolute disclosure risk methodology.

\subsection{Summary}\label{sec:analysis:summary}
Table \ref{tab:summary} summarizes our analysis. We abbreviate the methodology names: absolute disclosure risk with linking is \emph{AbsLink}, absolute disclosure risk without linking is \emph{Abs}, prior-to-posterior comparisons are \emph{Pr2P}, and the counterfactual methodology is \emph{Cf}. We use $\yes$ to denote that a methodology has a proof of concept with that particular desideratum, $\no$ to indicate that it is not supported at all, and $\maybe$ to indicate that it may be supported but a proof of concept is needed in order to demonstrate support---hence, $\maybe$ indicates potential areas of research focus. Note that it is not enough to support each desideratum with separate, incompatible proofs-of-concept; all the desiderata marked with $\yes$ for a particular methodology should be supported by a common proof-of-concept.
In general, we see that absolute disclosure risk methodologies have the fewest desirable properties if they are used for privacy/utility cost-benefit analysis. This is not to say that the posterior distribution that is analyzed by this methodology is not useful---this distribution is certainly informative, but only when used as part of a comparison as with the other two methodologies.

\begin{table}
    \centering
\begin{tabular}{|l|cccc|}\cline{2-5}
\multicolumn{1}{l|}{}& \textbf{AbsLink} & \textbf{Abs} & \textbf{Pr2P} & \textbf{Cf}\\\hline
\ref{sec:des:trans}: \textbf{Transparency} & \maybe & \maybe & \yes & \yes \\\hline
\ref{sec:des:universal}: \textbf{Universality} & \no &\yes & \yes & \yes\\\hline
\ref{sec:des:meta}: \textbf{Uninformative Statistics} & NA & \no & \yes & \yes\\\hline
\multicolumn{1}{|@{}l|}{\begin{tabular}{l}\ref{sec:des:arbitrary}: \textbf{Arbitrary Analyses}\\ (with remote access servers)
\end{tabular} }
& \maybe & \maybe & \maybe & \maybe\\\hline
\ref{sec:des:many}: \textbf{Arbitrarily Many Analyses} & \multicolumn{4}{c|}{Mathematically impossible}\\\hline
\ref{sec:des:recon}: \textbf{Avoiding Reconstruction} & \no & \yes &\yes &\yes\\\hline
\ref{sec:des:gen}: \textbf{Generalizable Inference} & \no & \no & \yes & \yes \\\hline
\ref{sec:des:composition}: \textbf{Composition} & \maybe & \maybe & \maybe & \yes\\\hline
\ref{sec:des:multiple}: \textbf{Multiple Attackers} & \no &\no &\yes &\yes\\\hline
\ref{sec:des:brittle}: \textbf{Resists Brittleness} & \yes & \yes & \yes & \yes\\\hline
\ref{sec:des:mis}: \textbf{Misspecification Resilience} &\maybe &\maybe &\yes &\yes\\\hline
\ref{sec:des:comp}: \textbf{\change{Control} Computational Effort} & \change{\no} & \change{\no} &\maybe &\maybe\\\hline
\ref{sec:des:manual}: \textbf{\change{Control} Manual Effort}  & \maybe & \maybe &\maybe &\maybe\\\hline
\end{tabular}
    \caption{Comparison of Risk Assessment Methodologies}
    \label{tab:summary}
\end{table}

\section{Discussion}\label{sec:discussion}
Confidential responses to the census have, in the past, been misused---for example, in World War II \cite{minkel}---and threats of abuse or misuse still exist.
The potential for misuse is widely believed to affect data quality, possibly contributing to the undercount of Hispanic populations in 2020 \cite{pesdemo}. Thus, a combination of access controls for the raw data and effective SDL to protect published statistics is necessary at many national statistical offices.

The efficient selection and implementation of a disclosure limitation system cannot be accomplished without a clear understanding of the intended uses for data. Independent of the degree of confidentiality protection (the privacy vs. utility trade-off), the choice of SDL approach (e.g., suppression, coarsening, or perturbation) will impact the flexibility of the resulting system to efficiently navigate that trade-off. The development of an \emph{optimal} disclosure limitation system also requires quantifying stakeholder utility for the resulting data products, quantifying the amount of disclosure risk in the data products, and acknowledging the mathematical fact that it is impossible to provide effective confidentiality protection while supporting every possible research purpose. (See further discussion of cost-benefit analysis in Appendix \ref{sec:si-costbenefit}.)

We consider three major methodologies for assessing disclosure risk: absolute disclosure risk, prior-to-posterior comparisons, and counterfactual comparisons. Of these three, absolute disclosure risk appears least useful for decision-making regarding algorithms for SDL, since it lacks important properties  for these risk assessments, summarized in Section \ref{sec:analysis:summary}. Continuing the analogy of \cite{hotzetal} that relates absolute disclosure risk to the chances of a patient dying, one would generally not make decisions about treatment options based on the overall probability of mortality, but rather on how the treatment would alter this probability---in other words, a comparison is needed. Thus, absolute disclosure risk is certainly an important quantity, but only useful for risk assessment when compared to something else, which is the purpose of the prior-to-posterior and counterfactual methodologies.

Hotz et al. \cite{hotzetal} criticized the disclosure avoidance system for the 2020 Census and called for a moratorium on using differential privacy to assess the privacy risk of data products. Some of their criticism gives a path for further research in differential privacy. 
For example, a 10x increase in an attacker's confidence about an attribute inference may be acceptable if the initial confidence was very small (e.g., $0.0001$) and less acceptable if the initial confidence were larger.
However, much of their criticism rests on the adherence of differential privacy to the mathematical limitation placed on \emph{all} disclosure limitation methodologies. This is not the fault of differential privacy, but rather a fundamental law \cite{Dinur2003}---some utility must be sacrificed to protect confidentiality, and all possible research uses cannot be simultaneously supported.  

Notably, none of the criticisms to date, (except \cite{christ:et:al:2022}, which compares non-optimized versions of swapping and differential privacy) have demonstrated any technique in sufficient detail to allow its privacy/utility trade-off to be compared to that of the 2020 DAS. Thus, following scientific traditions, researchers who say ``do X'' or ``don't do Y'' need to work out a proof of concept for a sufficiently rich application, defend it, and submit it to the scientific community for analysis of weaknesses. This is especially true for something as mathematically complex as disclosure risk assessment. Additional technical challenges that need to be resolved by absolute disclosure risk methodologies can be found in SI-4.

Data users and privacy advocates can legitimately debate the policy decisions that the Census Bureau makes regarding the proper balance of privacy vs. utility and regarding the prioritization of certain use cases over others for receiving greater accuracy in the published statistics. These are public policy decisions, and the Census Bureau should rely on the expertise of its diverse external stakeholders in making them. However, the technical tools such as algorithms and disclosure risk assessment methods should be based on the best available science. 
A call for a moratorium on the use of certain technologies should be replaced with an exhortation to use the best researched technology that is available at the time a disclosure avoidance system must be developed and deployed.

\section{Conclusions}\label{sec:conc}
\change{Transparent SDL techniques help promote the trust that is necessary for surveys based on self-response. Coupled with the importance of the 2020 Census data, their use }
%
has reinvigorated interest in disclosure limitation in federal statistics because the resulting data products affect countless individuals via public policy and research.
By laying out principles for the analysis of disclosure risk assessment methodologies, the aim of this paper is to help frame the debate on how the federal statistical system should proceed, and to identify the necessary research directions.

\newpage
\appendix
\section*{Appendices}









In these supplementary materials we provide a discussion of the following items:
\begin{enumerate}

\item[A.] Details of how the U.S. Census Bureau engaged with stakeholders to identify major use cases for census data and assessed how differing degrees of SDL protection would impact the utility of data to support those use cases.

\item[B.] A comparison of the flexibility afforded by different SDL strategies when designing disclosure avoidance systems.

\item[C.] A discussion of synthetic data, a relevant disclosure avoidance method that did not get complete coverage in the main manuscript. 

\item[D.] A discussion of cost-benefit and social welfare analysis under different disclosure risk methodologies.

\item[E.] Errors in critiques of the Census Bureau’s risk analysis and differential privacy implementation for the 2020 Census.

\item[F.] Documentation of the statement (referenced in  the main text of this paper) by a state demographer regarding reconstruction of microdata from published census tabulations.

\item[G.] Copy of personal communication from Gong and Meng.

\end{enumerate}

\section{Quantifying utility via stakeholder engagement}
\label{sec:si-utility}
 
To ensure representation of a broad spectrum of users, the agency met extensively with external stakeholder groups, including federal advisory committees;  American Indian and Alaska Native tribal leaders; federal, state, and local officials; state and local demographers on the Federal State Cooperative for Population Estimates; members of the Census Bureau's State Data Center and Census Information Center networks; civil rights groups; privacy advocates; and many more. In addition to soliciting formal public comment in the Federal Register \cite{FRN:2018}, the agency catalogued its efforts continuously \cite{USCB:DAmodernization:2022}.

Through these consultations, the agency developed the Detailed Summary Metrics \cite{dsm}, which included  measures of absolute error, relative error, and outliers for a wide array of tabulations at varying levels of geography and for geographic entities of differing population sizes.
For example, Wright and Irimata \cite{wright_irimata}  studied the impact on redistricting and Voting Rights Act use cases. They proposed utility measures for ensuring that the largest demographic group within small places and minor civil divisions (a proxy for arbitrary, contiguous legally-defined geographic areas) with sufficient population to constitute an individual voting district, as a proportion of that geographic area's total population, was within 5\% of the enumerated value at least 95\% of the time.

To obtain additional feedback and identify use-cases or problems not adequately covered by the detailed summary metrics, the agency  released eight sets of demonstration data products between 2019 and 2022, generated by running 2010 Census data through successive versions of the 2020 DAS \cite{ddp}.

This  unprecedented level of public engagement for a disclosure avoidance system informed decisions about many design choices, such as the algorithm's geographic processing hierarchy, noise distributions, privacy budget allocations for different geographies and population characteristics, etc. Feedback from American Indian and Alaska Native tribal leaders about the absolute and relative error in statistics for small tribal areas, for example, led to the incorporation of a tribal area branch of the algorithm's geographic processing hierarchy, which is the method used by the TopDown algorithm in the 2020 Census Disclosure Avoidance System to process the census data beginning with the nation as a whole, then proceeding to states, counties, census tracts, block groups, and census blocks. See \cite{tdahdsr} for details.

\section{A comparison of the algorithmic flexibility of SDL frameworks}
\label{sec:si-flexibility}

Statistical agencies routinely use a risk management framework to determine when disclosure risk has been sufficiently mitigated, balancing the residual disclosure risk against the societal value of releasing the data product. Central to this assessment is the resulting utility of the data after SDL has been applied. All SDL techniques impact data utility. Whether by removing data (suppression methods), reducing precision (coarsening methods), or by otherwise introducing uncertainty (so-called perturbative methods), SDL methods reduce disclosure risk by reducing the availability and/or the precision of the data to be released. The selection of one SDL method over another or one set of implementation parameters over another has profound implications for the suitability of the resulting data for particular types of analysis.  For example, income can be a highly disclosive data element because many data subjects' exact incomes make them population uniques. If an agency decided to protect income by coarsening individuals' income data into ranges for a public report,  then the ranges
that would support an analysis of extreme poverty can be very different from the ranges that would be  used to assess gender disparities in CEO compensation. There are many other use cases of income data, and the agency would need to decide on a set of ranges that would balance the trade-off between these applications. Alternatively, the agency could permit some analyses on the unaltered confidential data and apply SDL to the outputs, for example by limiting the published precision of model parameters or infusing noise into them. Either way, the agency must assess the incremental disclosure risk of the released statistical data.


In order to support many heterogeneous use cases, it is important to choose SDL frameworks that provide the greatest flexibility and to recognize which frameworks and implementing mechanisms would limit the ability of the agency to meet its goals. When considering SDL frameworks for the 2020 Census, the Census Bureau strongly weighed the data utility implications of the leading SDL methods: suppression (item, cell, or table), coarsening, data swapping (often followed by aggregation), and noise infusion. 

Item, cell and table suppression have  often been used to protect tabular data \cite{sdcpractice}. Items, cells or whole tables with small counts or magnitudes, as appropriate, are initially suppressed because they are deemed too disclosive, usually called \emph{sensitive} in this context. However, because tabulations often overlap, it would be possible to recover those suppressed data by adding together and subtracting elements from other tables. This is called a \emph{differencing attack}, one example of a database reconstruction attack. Thus, a second round of suppression, called \emph{complementary suppression}, is also needed to completely mask the data in the sensitive cells. The combination of primary and complementary suppression is known to cause significant loss of data and was a criticism of disclosure avoidance systems in prior censuses \cite{mckenna:2018}. Abowd and Schmutte \cite{abowd_schumtte:2015} pointed out that primary/complementary suppression is inherently nonignorable because the selection criteria depend on the values of data in the suppressed cells. Furthermore, Chetty et al. \cite{Chetty_Friedman_2019} found that suppression was significantly outperformed by techniques inspired by differential privacy for the Opportunity Atlas. The root cause of these issues is that suppression  does not provide much flexibility---a cell is either suppressed or not. In contrast, noise infusion provides a continuum between perfectly precise cell counts and completely masked cell counts. 

Coarsening of variables on individual records presents difficulties similar to suppression. If all records are coarsened in the same way, significant information is lost. If records are coarsened in different ways, then aggregating records becomes problematic.

Data swapping is a noise infusion technique in which pairs of individuals, or pairs of households, are matched on some variables (called the sort key) and other attributes are interchanged between the matched records. For example, the match key might be total household size and the exchanged attribute might be the geographic identifier. This swap effectively puts the data from one record into the geography of the other.  
Swapping introduces some uncertainty into the published data but, because it is a record-level perturbation method, it does not provide a good privacy/utility trade-off. For example, consider a set of microdata that has been swapped at a rate of 10\% (so one in ten records has been altered). Given any record,  attackers would have at least 90\% confidence that they can correctly identify sensitive attributes of the corresponding individuals. Additional knowledge about an individual may even be used to reverse some of the swapping. However, a 10\% swap rate can also have a significant impact on utility and so an agency would prefer to choose a focused approach in which swapping is mostly concentrated on individuals who have a higher risk of disclosure, typically those that are considered to be ``population uniques.'' For highly granular data collections, however, this can be difficult as the number of population uniques typically increases with the granularity of the data publication.  For example, in 2010, 57\% of the person records in the decennial census were unique based on their age, sex, block, race, and ethnicity characteristics \cite[p. 4]{abowd2_fairlines_appendixB}. That percentage would be expected to increase significantly when uniqueness of household composition is considered. Thus, an uncomfortably large swap rate would be needed to provide meaningful levels of privacy protections if swapping is the only disclosure avoidance method being used.

Although data swapping was approved for use in the 2010 Census, the agency believed that the swap rate was insufficient to protect confidentiality without further measures. Those additional measures included coarsening, aggregation, and synthetic microdata, which were used for the tabular products in the belief that in combination with swapping they would provide sufficient protection. However, because the 2010 data products included a large number of detailed tabulations, it is often possible to undo the aggregation, especially in census blocks with low populations. 
The Census Bureau withheld publication of a detailed analysis of the vulnerabilities until after the 2020 Census was completed and the full 2020 disclosure avoidance system could be tested for similar vulnerabilities; however, the methods were rigorously peer reviewed \cite{JASON:2020,JASON:2022}. 

The use of data swapping for future censuses would require significantly larger swap rates and less detail in the tabulations that are produced. Such a system has many fewer tuning possibilities than one based on formal privacy methods.


This left noise infusion for aggregate statistics as the Census Bureau's leading SDL candidate framework for the 2020 Census. With noise infusion, one could choose the noise distribution, the scale (amount) of noise, and the statistics to infuse with noise. For example, one isn't limited to ``all-or-nothing'' perturbations as in suppression and coarsening. One also is not forced to work at the record level as in data swapping (which is responsible for its problematic privacy/utility trade-off). Hence noise infusion provides a great deal of flexibility and tuning parameters for the 2020 disclosure avoidance system.

\section{Disclosure limitation using synthetic data}
\label{sec:si-synthetic}
Synthetic data is an orthogonal approach to noise infusion or suppression.  Any disclosure avoidance method, including synthetic data, will impact data utility for certain analyses in order to provide some confidentiality protection.  Synthetic data still carry disclosure risks, as do all data protected using any disclosure avoidance method. The key question regarding the use of synthetic data is not how it compares to unfettered access to the confidential data for anyone who wants it---that is not possible for many data sets, at least not under current laws and societal demands for data privacy---but how it compares against other disclosure avoidance frameworks.  Agencies and other data custodians need to ask themselves bluntly: what are the alternatives for the confidential data at hand?  In this section we describe the broad use of synthetic data to support the dissemination of data products derived from confidential data and provide a risk-utility discussion.

Several agencies now use or consider synthetic data  frameworks when creating public-use files from confidential data.
As examples, the Census Bureau---which already has released synthetic public-use files for the Survey of Income and Program Participation  \cite{abowd06} and Longitudinal Business Database  \cite{lbdisr}---is researching significant changes to the confidentiality protection methods for the American Community Survey, its flagship survey, to enable the use of synthetic data \cite{acssyn} and validation servers. The Internal Revenue Service is making a synthetic public-use file of individual tax returns \cite{syntheticirs}.  The Agency for Healthcare Research and Quality has funded a project to create a synthetic version of healthcare claims data \cite{ahrqsynthetic}. Additionally, as described in a manual on synthetic data for national statistics  organizations  recently authored by the United Nations Economic Commission for Europe \cite{unecesynthetic}, synthetic data have been used by Statistics Canada, Statistics New Zealand, the Australian Bureau of Statistics, the United Kingdom Office of National Statistics, and in the Scottish Longitudinal Study, to name a few others.  The private sector also recognizes the appeal of synthetic data approaches: there are now dozens of companies advertising their abilities to create synthetic data solutions.

The Longitudinal Business Database (LBD), which contains the annual total payroll and employee size since 1975 for every
U.S. business establishment with paid employees, is an informative case study on how synthetic data can enable public access to otherwise confidential data \cite{lbdisr}.  Data from the LBD derive from Census Bureau surveys and tax files compiled by the Internal Revenue Service (IRS), making the LBD subject to the confidentiality protection provisions in both Title 13 and Title 26  of the U.S.\ Code.  As a result, no actual values for individual establishments in the LBD can be released to the public; even the fact that an establishment filed taxes---and hence is in the data set---is protected.  Thus,  top coding cannot be used on monetary variables as a large fraction of exact values would be released.  This also suggests that swapping would have to be done at an extremely high rate, in which case the released data would be useless for any analysis involving relationships with swapped variables. Swapping also would not protect Title 13 confidentiality because so many of the magnitude variables are population uniques and other agencies, in particular the IRS, also have copies of the data.  Furthermore, many variables of interest to
researchers and policy makers, for example the number of employees and total payroll, have skewed distributions with thick tails even within industry
classifications. For simple noise-infusion strategies, the amount of added noise necessary to disguise these observations  would have to be very large, also resulting in
data of limited usefulness.  Among currently available alternatives, the only way that the Census Bureau and IRS were willing to release a public-use version of the LBD was to make it fully synthetic \cite{lbdisr}.

Any researcher can use the synthetic LBD---which currently includes synthetic establishments in existence some time between 1975 and 2011---from a dedicated Census Bureau website \cite{synLBD}. After refining the analysis and modeling strategies based on explorations with the synthetic data, the researcher can request that the Census Bureau run the analysis on the confidential data. The Census Bureau provides results of this confidential-data analysis, processed by SDL on the outputs of the analysis to manage the disclosure risks.

Another illustrative example is the Survey of Income and Program Participation (SIPP) synthetic data program \cite{synSIPP}. In this program, the Census Bureau first combines survey responses from SIPP panels with administrative tax and benefits data into a gold standard file (GSF).  The integration of these sensitive values and SIPP variables creates serious risks of disclosure and consequent harm to SIPP participants. Hence, to make a public-use version of the GSF, the Census Bureau turned to a synthetic data solution.  It uses sequential regression techniques to synthesize nearly every variable---a handful of variables remain at their collected values---for every respondent record in the GSF \cite{benedetto:et:al:2018}.  Among existing alternatives, this was the only disclosure limitation method trusted by the providers of the administrative data to protect respondents' identities and attributes, especially against linking to other available SIPP products such as the previously released SIPP public-use microdata files.  

To use the synthetic version of the GSF, users submit a request for access, after which they can perform any and all analyses supported by the schema of the GSF. 
The SSB program also offers validation of results on the GSF. Users provide their analysis code, proven to run on the synthetic GSF, to Census Bureau staff, who then execute the code on the GSF itself and return the results.  As with the LBD, the results are processed by output SDL.  

As a final example, synthetic data methods have been used to protect the confidentiality of respondents in group quarters (GQ) facilities in the American Community Survey  since 2007, and in the 2010 Census \cite{mckenna:2019:microdata}. GQ respondents can present difficulties for other disclosure avoidance methods that depend on matching records, such as data swapping, as the characteristics of GQ respondents can vary significantly among GQ types. For example, swapping data values of individuals in a men's prison with those in a women's shelter likely would result in data of dubious quality.
Synthetic data methods can flexibly accommodate GQ data, in that the typical frameworks used for synthetic data, such as sequential modeling, can accommodate models that apply to single GQ facilities or that apply to the same type of GQ across states.

As these implementations illustrate, synthetic data systems have provided researchers with access to  sensitive microdata in ways that would not have been possible with the other  SDL frameworks that were in existence.   We next explain why synthetic data solutions were used by considering some alternative disclosure avoidance methods traditionally employed by federal agencies and other data custodians.   These alternatives and the corresponding methods for assessing their disclosure risks generally do not satisfy many of the desiderata described in the main text and in Section 2 of this supplementary information; however, they represent hypothetical alternatives that one might consider when restricted to traditional disclosure avoidance methods and averse to the use of synthetic data.

{\bf Alternative 1: No access to public-use microdata files.}  One alternative is to deny access to public-use microdata sets, except possibly to vetted and approved users via physical or virtual data enclaves. While limiting access can manage disclosure risks, it sacrifices some of the societal benefits of broad access to microdata.  In particular, it would serve as a barrier to many user groups.  For example, there are 33 Federal Statistical Research Data Centers (FSRDCs) across the United States.  Many U.S.\ residents do not live close enough to FSRDCs to take advantage of the access they provide, although there is now a virtual access option that more than 400 researchers currently use. Because the FSRDCs primarily serve the research community (by design), access restrictions could exclude broad swathes of the public, such as local entrepreneurs, government planners, students learning the skills of data science as applied to public policy and the social sciences, and citizen scientists seeking to learn about their communities. Such groups represent legitimate stakeholders in the benefits of agency data. Accessing these data in a privacy-protected way serves that interest.  Thus, while we believe providing vetted and approved users and projects access to confidential data should be a component of agencies' data dissemination policies,  we expect the reduction in social welfare from Alternative 1 would be too great to make it a preferred strategy for data access, although it has long been the practice of agencies for business data dissemination.

{\bf Alternative 2: Access to microdata treated with traditional disclosure avoidance applied at low intensity.}  For many agencies, past and current practice is to perturb only a small amount of data, with the hope of discouraging intruders while disturbing relationships and distributions in the data as little as possible.  For example, swap only a low fraction of records, e.g., $<1\%$; suppress only a small number of values in the data; or coarsen only slightly. Unfortunately, as noted in \cite{hotzetal} and repeatedly in the research literature on disclosure avoidance, low intensity traditional disclosure avoidance methods may simply no longer be relied upon to protect confidentiality \cite{abowd_schumtte:2015, dworkarsa, kellershipp}.  For example, consider a data set with dozens of variables, including many demographic characteristics readily available in external files.  As discussed in the main text, with enough variables everyone is essentially a population unique. Since swapping 1\% of records means that 99\% of records are left untouched, privacy attackers may not feel discouraged from linking to external files.  As another example, collapsing geography to high levels, say the public-use microdata areas (PUMA) used by the Census Bureau for the American Community Survey, may not matter when a data subject's combination of characteristics still makes them unique within that PUMA.  Thus, if agencies place a high value on privacy, Alternative 2 falls short.

Incidentally, light touch disclosure avoidance methods do not necessarily have light impacts on data utility.  As examples, even low rates of data swapping can deteriorate confidence interval coverage substantially \cite{drechsler:reiter}, and top-coding of the upper tails of income can distort measures of income inequality \cite{kennickel:lane}.   

{\bf Alternative 3: Access to microdata treated with traditional disclosure avoidance applied at high intensity.}  A third option is to apply traditional disclosure limitation methods with high intensity, with the goal of achieving better privacy protection.  For example, agencies could suppress very high fractions of values; perform swapping across all  records and variables; or add noise with huge variances. We conjecture that data users would find such data far less useful analytically than synthetic data,  where records are generated from models that preserve important relationships in the data. And there is support for our conjecture from the Census Bureau's own experiments with high-intensity swapping procedures \cite[Table 4]{abowd:hawes:2023}.

{\bf General comments on disclosure risks and data usefulness.}  We close this section with additional comments on the disclosure risks and the data utility associated with synthetic data.  

Regarding disclosure risks, it is important to assess disclosure risks before releasing synthetic data, as they do not guarantee zero risk \cite{hotzetal}. Disclosure risk checks were done before the release of the synthetic LBD and the synthetic SIPP data products using best known approaches at the time.  For the synthetic LBD, these checks focused on attribute disclosure risk assessments, e.g., how accurately could an intruder learn the payroll of large establishments by using the synthetic data?   For the synthetic SIPP data, the checks emphasized the success of record linkage of the synthetic records to the confidential records.  The insights from differential privacy tell us that such checks, while informative, do not convey mathematical guarantees of privacy against broad classes of attackers.  Hence, as suggested in \cite{hotzetal}, additional research is needed on assessing disclosure risks in complex synthetic data, as well as on generating synthetic data with formal privacy guarantees.

On the other hand, synthetic data sampled from stochastic models have an appealing privacy feature not present in low-intensity traditional disclosure avoidance strategies. In practical applications, the randomness in drawing synthetic values, when implemented appropriately (so that the model does not overfit and memorize some of the training data),  can strengthen privacy protections, and can even be differentially private under certain conditions \cite{wang:et:al:2015}. Intuitively, this is because, with high probability, this process creates records where the synthetic values do not exactly match the multivariate set of confidential data values.  In contrast, swapping at low rates leaves nearly all released data at the confidential values, as does coarsening applied to only a small number of variables.      

Regarding data utility, synthetic data have a well-known limitation: whatever is in the synthesis models is passed on to the analysis. Undoubtedly, some analyses will not be preserved.  However, one should view the utility of synthetic data not against using the confidential data, but against other disclosure avoidance methods that provide protection akin to synthetic data, such as a 100\% data swap (in the case of fully synthetic data) or massive suppression or coarsening of all variables.  As noted previously, high-intensity applications of these methods can seriously degrade the utility of the data relative to synthetic data, which has the potential to preserve distributional features and relationships across variables. For the LBD and SIPP examples above, the synthetic data support very high dimensional analyses, similar to public-use microdata and much broader than the two- and three-way tabular summaries published using traditional disclosure avoidance.

It is also important to emphasize that many analyses of public-use microdata, indeed perhaps the majority for some agencies' data products, involve relatively low dimensional summaries, such as estimates of means or counts by groups.  Experience shows that  synthesizers can enable accurate estimation for many low-dimensional analyses  \cite{lbdisr, abowd06}.  That said, in practical applications of synthetic data approaches, it is not ``all or nothing.''  Researchers whose sophisticated models are not likely to be preserved, such as the examples identified in \cite{hotzetal}, can gain access to the confidential data under restricted-use agreements, such as via FSRDCs.

\section{Cost-benefit and social welfare analysis under different disclosure risk methodologies}
\label{sec:si-costbenefit}
Hotz et al. \cite{hotzetal} suggested that cost-benefit analysis should be applied to managing the utility-disclosure risk decision-making without offering examples or implementation suggestions beyond their preferred disclosure risk methodology. We consider the nature of cost-benefit analysis under different approaches to disclosure risk assessment. More specifically, assuming a given data publication task and a disclosure risk methodology, how would one use that disclosure risk methodology to guide publication? In this context, the data provider needs to choose \emph{SDL framework and implementing mechanisms} that describe how the publication task will be carried out. For example, an SDL framework and mechanism might be, say, ``cell suppression under a given p-percent rule'' or ``swapping on a given set of characteristics and with known swap rate.'' In the context of formal privacy, it might be ``$\varepsilon$-differential privacy implemented with the Laplace mechanism at a known $\varepsilon$''.

The cost-benefit analysis described by \cite{hotzetal} suggests that the right approach is to rank different SDL frameworks and associated implementation mechanisms in terms of their net benefits across all members of society and then choose the best one. There are two basic models for this task. First, if everyone has the same ranking over SDL frameworks and implementations, then the basic task is to assess the common ranking of these SDL regimes and choose the best one: the \emph{universally optimal SDL regime}. Second, if there are heterogeneous preferences over SDL regimes, then it is necessary to devise a social welfare criterion that balances the competing interests of different parties, in the spirit of Abowd and Schmutte \cite{abowdschmutte2}.

For this analysis, we assume the costs in question are the disclosure risks faced by the data subjects. The benefits are the improvements in generalizable inferences about the population from which the data are drawn. For shorthand, we refer to the latter as \emph{data quality}, with the caveat that, in general, different users can have different assessments of the quality of a given publication depending on their background information and use case. In principle, the cost-benefit analysis might incorporate other desiderata, such as the direct monetary costs of implementing different regimes or the effects of different regimes on response rates. We note that conducting this broader cost-benefit analysis would require the ability to trade off disclosure risk against data quality in our more limited sense, making the simplified analysis a useful starting point. If a disclosure risk methodology makes our limited cost-benefit analysis impossible, it is obviously not a candidate for the broader social choice problem. 

\subsection{Universally optimal SDL regimes}

For the statistical agency to select a universally optimal SDL regime, it must be able to unambiguously rank all regimes such that all data users and data subjects share that same ranking. 
Under the absolute disclosure risk criterion, selecting a universally optimal SDL regime requires extremely restrictive assumptions. 
In general, it requires that all attackers have the same prior distributions over the sensitive information of each individual. Otherwise, we can not meaningfully rank the costs to individuals associated with different regimes. An alternative might be, as in some implementations of the prior-posterior methodology, to estimate the absolute disclosure risk assuming the worst-case prior. Then, because the likelihood component of the posterior distribution is fully determined by the SDL mechanism, the absolute disclosure risk would satisfy the homogeneity condition required to unambiguously rank all regimes.

\subsection{General social welfare analysis}

Following \cite{abowdschmutte2}, it is possible to accommodate heterogeneity in costs associated with disclosure risk and in the benefits associated with data quality. With such heterogeneity, there is generally no universally optimal regime, but as long as a social welfare function can be specified, the problem of finding its maximum value is well-specified. The main requirements are that (1) it is possible to define social welfare in terms of the data quality measure and the disclosure risk measure, and (2) each SDL regime can be associated with a data quality measure and disclosure risk measures. The problem of maximizing the social welfare function may not always have a solution, but when it does, it is achieved by balancing the marginal cost of increasing disclosure risk with the marginal benefit of increasing data quality. Even if the social welfare function is stated in terms of absolute disclosure risk, and summarizes heterogeneous preferences over that risk, the technology need not be based on absolute disclosure risk. The relevant technological constraint is the solution to the production possibilities frontier, which can be set up to accommodate parameterization by absolute, counterfactual, or prior-posterior disclosure risk. The parameters defining the best feasible trade-off between disclosure risk and data quality also enter the social welfare function, where the disclosure risk parameters are converted to the desired metric by applying each person's prior before aggregation, just as each person's data quality valuation is applied before aggregation. 

We do not mean to imply that the general social welfare analysis is feasible to implement, but it can guide policy making. For example, in setting parameters for the 2020 Census DAS, the Data Stewardship Executive Policy committee consistently allocated the minimum privacy-loss budget sufficient to meet stated data quality metrics specified in support of known, leading use cases, as described in the main text and \cite{abowd:hawes:2023}. This can be interpreted as behaving as if the production possibilities frontier were the one described by the mechanisms implemented in the 2020 DAS and the social welfare function had rectilinear iso-utility curves over the privacy-loss budget and the data quality measure.

\section{Details on erroneous analyses of the Census Bureau reconstruction experiment and implementation of differential privacy}
\label{sec:si-errors}
\subsection{Errors in published critiques}

In the disclosure avoidance system for the 2010 Census, confidentiality protections for the tabular data were provided mainly by the following techniques \cite{mckenna:2018,sf1:2010:tech}:
\begin{itemize}
\item Data swapping for the population in households.
\item Partially synthetic data for the population in group quarters.
\item Age coarsening for tables at the block level and race coarsening for tables at the tract and county level.
\item Aggregation of records in the universe into summary tables.
\end{itemize}

Early results from the Census Bureau's reconstruction-abetted reidentification attack were reported in the course of litigation seeking to prevent the use of differential privacy by the Census Bureau \cite{abowd1_fairlines, abowd2_fairlines_appendixB, abowd3_fairlines}. These results showed that in an earlier version of the reconstructed data, it is possible to obtain exact-age matches to 46.48\% of the records in the 2010 Census Edited File (the confidential version of the 2010 Census data) based on the attributes block, sex, exact age, race, and ethnicity. The percentage was significantly higher using approximate matching based on age within $\pm 1$ or age binned to 38 categories supported by the block-level tables in Summary File 1. Agreement was even higher when the analysis focused on low-population blocks \cite[Table 1]{abowd3_fairlines}--these are the geographies for which reconstruction is expected to work better. They are also the types of geographies that are widely considered to have higher disclosure risk because of the prevalence of population uniques on three widely available pseudo-identifiers: census block, sex, and age. The reconstruction portion of the Census Bureau's simulated attack only defeats aggregation and does not attempt to undo the household address swapping operations.

Meanwhile, several analyses \cite{rvr,Muralidhar:2022,muralidhar:domingo-ferrer:2023,Francis2022,kenny:et:al:2021,kenny:et:al:2022} asked related interesting questions. 

Ruggles and Van Riper \cite{rvr} ask: is 46.48\% really high? To answer this question, they considered how well an attacker might do by random guessing. They concluded that a random guesser would actually have a higher success rate than 46\%. However, their methodology was severely flawed. Using their methodology, the highest success rate (approximately $57\%$) would be achieved by an attacker who guesses that everyone is a 50-year-old female \cite[pp. 15--18]{abowd3_fairlines}, even though 50-year-old females were less than $1\%$ of the 2010 population. If 50-year old females are 1\% of the population, then once one has reidentified that 1\%, the job is finished. The 57\% ``reidentification rate'' the Ruggles-Van Riper algorithm estimates is actually the probability of finding a 50-year old female in a random block, or more generally, any output from their algorithm estimates the probability of finding a random person with a given sex and age in a random block. That is not a reidentification rate.

Francis \cite{Francis2022} asks: can a simple baseline outperform the attribute inference (identification of the pair $\{race$, $ethnicity\}$) that was conducted with the aid of the reconstructed Hundred-percent Detail File (HDF)? However, the methodology used for reporting early results to assess the reconstruction experiment differed from the methodology Francis used to assess his results. First, his work takes the Census Bureau's results, which are grouped by block population size, and compares them to his own results, which are grouped by racial homogeneity. The resulting group-by-group comparisons are uninformative because of this misalignment. They do not allow drawing any proper conclusions. His results are also affected by whole-person census imputations (where actual characteristics are not known, but copied from another member of the household or a neighbor) and do not account for the homogeneity in races that is introduced by these edits and the data swapping keys.

Muralidhar \cite{Muralidhar:2022} asks: is reconstruction really as easy as the Census Bureau experiments suggest? To answer this question, he makes a small-scale reconstruction that excluded important tabulations, used an incorrect and much simpler schema, and incorrectly asserted that the existence of multiple solutions with respect to some schema was sufficient to ignore reidentification results. With his co-author, Domingo-Ferrer, \cite{muralidhar:domingo-ferrer:2023} they assert that the reconstruction does not imply reidentification because of suppressions that accompany correctly-implemented disclosure limitation; however, the 2010 Census data published in Summary File 1 did not impose suppression, making this argument false \cite{garfinkel:2023}. 

Kenny et al. \cite{kenny:et:al:2021} ask: is high quality inference through statistical means the same as disclosure via reconstruction? They conclude that it is. This is an incorrect conclusion because disclosure avoidance must protect how an individual's data differ from statistical predictions about them. Statistically speaking, it is the distinction between the data-generating distribution and the actual realized values of the random variables. It is also the difference between leave-one-out (LOO) error estimates of a model's accuracy compared to its in-sample training set accuracy. 

In the rest of this supplementary section, we describe the flaws in these methodologies in more detail. We explain how to confirm the errors of Ruggles and Van Riper using their simulation code \cite{rvrcode}. We provide more information about the discrepancies with the baseline provided by Francis. We show that Muralidhar's analyses gives results similar to the Census Bureau's when the mathematics are corrected. Finally, we show that Kenny et al. confound leave-one-out generalizable inference with non-leave-one-out privacy violating inferences.   
Kenny et al. also use flawed statistics that fundamentally overstate the error in the Census Bureau's differentially private disclosure avoidance system for the 2020 Census. 

\subsection{Methodological errors in Ruggles and Van Riper and their simulation}
\label{sec:ruggles_errors}

When records from the published SF1 tables are matched to the CEF in the Census Bureau's or other reconstruction experiments, the matching must be one-to-one as described in the appendices to legal declarations \cite{abowd2_fairlines_appendixB}. A record  in the reconstructed data may be assigned to at most one distinct record from the confidential data. For example, suppose the reconstructed data contain the records $\{A, A, B, C\}$ and the confidential data contain the records $\{A, B, D, D\}$. There would be exactly two matches: one of the $A$ records from the reconstructed data would be assigned to the $A$ record of the confidential data, and the $B$ record in reconstructed data would be matched to the $B$ record in the confidential data.

The comparison methodology developed by Ruggles and Van Riper assesses the success rate of an attacker in a different way \cite{rvr,rvrcode}--there is no attempt to perform one-to-one matching between an attacker's file and the ground truth file (which would correspond to the confidential data). Instead, a block is chosen using population-weighted random sampling. If one block has $1,000$ people and a second block has only $1$ person, the first block will be $1,000$ times more likely to be selected. After a block is selected, the attacker guesses a combination of sex and age. If this combination matches \emph{anyone} in the block then the attacker is credited with a success. The evaluation metric of \cite{rvr,rvrcode} is the probability the attacker succeeds. 

In this methodology, if a block is chosen 10 times, the attacker makes 10 guesses for the block. If the block has only one $50$-year old female, but the attacker guesses $50$-year old female each time, the attacker is credited with 10 matches. In other words, the matching is not one-to-one as 10 records from the attacker file are matched with just a single record in the population ground truth.

Since large blocks are selected more often and since they usually have a ``$50$-year old female'' (it was the most common combination of age and sex in 2010 Census), the attacker should always guess this combination to maximize their success probability. This attacker would be credited with a $57\%$ match rate in the methodology of \cite{rvr,rvrcode} even though $50$-year old females were less than $1\%$ of the entire population.

The $57\%$ estimated by Ruggles and Van Riper is the simulation estimate of the population-weighted probability that a random block contains a $50$-year old female. In other words, Ruggles and Van Riper are not measuring how well an attacker's guess matches the demographics of the population, they are  only measuring how likely they are to find a random age-sex combination in a population-weighted randomly chosen block. 

\subsection{A Guide to the Ruggles and Van Riper simulation}

This flaw can be demonstrated in a repeatable way using the provided code \cite{rvrcode}. First, download three files from their replication archive:
\begin{itemize}
\item censim3.f: the Fortran code that runs the simulation.
\item agesex.txt: a data file containing the cumulative distribution histogram of the age and sex of the simulated population.
\item blocksize3.csv: a data file containing the cumulative distribution of block population sizes.
\end{itemize}
Next, change lines 11 and 12 in censim3.f so that they refer to the locations of the two downloaded data files. In the Fortran file, lines 62-65 implement the guess of the attacker. By adding the line \verb|hyp=146| after line 65, one forces the attacker to always guess just one age-sex combination (corresponding to 50-year-old females). Compiling and running this code results in a file called outfile.txt that tracks the successes of an attacker. It is a spreadsheet with multiple columns. The first is the block id, the second is the population of the block, the third is how many of the attacker's guesses for the block matched \emph{someone} from the block (the attacker guesses could match the same person multiple times, contributing to this count each time). The fourth column is the number of attacker guesses for the block that failed to match \emph{anyone} in it. The success rate from the methodology of \cite{rvr} is the sum of the third column divided by the sums of the third and fourth column. Note that each run of the code may produce slightly different results, because each run creates a random synthetic population that serves as the ground truth in the simulation.

\subsection{Methodological errors in the analysis of Francis} 
\label{sec:francis_errors}

Francis \cite{Francis2022} proposed a simple comparison baseline for predicting an individual's race and ethnicity pair from the published tabulations: if the individual's block contains 5 or more people, guess the modal (most common) race/ethnicity in that block. First, note that this baseline would have no accuracy for people who differ from the modal race/ethnicity combination, but the Census Bureau's reconstruction experiment has high accuracy for these sub-populations as shown \cite[Table 6]{abowd2_fairlines_appendixB} and [Table 3]\cite{abowd:hawes:2023}.

The main finding by Francis is that 11\% of the United States population in 2010 appeared to live in blocks that contained at least 5 people and in which the race and ethnicity were all the same. Thus, naively, one would be expected to identify the race and ethnicity of 11\% of the population with complete confidence. However, the 2010 data had nearly 6 million whole-person imputations \cite{coverage2010} and data swapping served to make blocks appear even more homogeneous \cite{mckenna:2019:microdata}. Without adjusting for these factors, Francis's analysis is overly optimistic. 

In contrast,  the Census Bureau's reconstruction experiment treated race and ethnicity as correctly inferred only when there was a record in an attacker's dataset and the record, augmented with information from the reconstructed data,  matched a ``data-defined person'' (i.e., not a product of whole-person imputation) in the unswapped, confidential CEF. All truly homogeneous blocks would have perfect precision in the baseline of Francis and also in the Census Bureau's experiment. However, the Census Bureau used the narrower data-defined population to avoid allowing imputations and swapping to inflate the results.

Furthermore, the Census Bureau results were tabulated by block size, whereas the experiments of Francis were tabulated by block homogeneity (the fraction of individuals in a block that share its most frequent race $\times$ ethnicity combination), making his direct comparisons inapplicable. The Census Bureau 
reconstruction experiment has very high accuracy for minority populations in non-homogeneous blocks \cite[Tables 2 and 3]{abowd:hawes:2023}. For the same sub-populations, the baseline from Francis would achieve a precision of exactly zero.

\subsection{Methodological errors in Muralidhar's analysis}
\label{sec:muralidhar_errors}

Muralihdar performed a simple example of database reconstruction using a single tract from the 2010 Census. He analyzed variation in the space of its feasible solutions. We agree that this type of analysis has merit. We do not agree that the existence of positive solution variability directly undermines interpretation of reidentification statistics. The attacker using reconstructed microdata as part of a record-linking reidentification algorithm does not have access to the sensitive data, so if solution variability were an obstacle to reidentification, then reidentification rates should be depressed by solution variability. Moreover, solution variability can be upper bounded using public data. When that upper bound is zero, as it is for about two-thirds of 2010 Census blocks \cite[Slide 17]{hawes:2022}, then, an attacker can explicitly focus attacks on areas where they are certain that aggregation offers no additional confidentiality protection and reidentify nonmodal persons with 95\% confidence \cite[Slide 19]{hawes:2022}. Here we elaborate these arguments while showing that Muralidhar used an incorrect version of the 2010 Census data schema, selected an unrepresentative block to study, and incorrectly concluded that the 2010 data schema used in our reconstruction was too coarse to permit reidentification.

Reconstructing microdata from aggregated tabular data releases is accomplished by converting the published tables into a system of integer-valued linear equations, then solving for the set (or sets) of individual-level records that conform to the constraints imposed by those equations. The more source tables that are included in the reconstruction, the greater the number of constraints that a feasible microdata reconstruction must satisfy. Muralidhar \cite{Muralidhar:2022} attempts a single-tract example of the reconstruction of 2010 Census microdata in order to assert that the results of the Census Bureau's simulated reconstruction-abetted reidentification are less concerning than the Census Bureau's Data Stewardship Executive Policy (DSEP) committee considers them to be. In doing so, he asserts that the reconstruction process typically yields a ``very large'' number of possible solutions---different sets of reconstructed microdata records that all satisfy the established constraints. His example involves the reconstruction of microdata records for a single census tract (5.01) in Laramie County, Wyoming. 

As described  briefly in \cite{hawes:2022}, the Census Bureau did compute solution variability upper bounds for all blocks in the 2010 Census using the 38-bin age coarsening supported by the block-level tables. This solution variability for the block-level-schema upper-bounds tract-level solution variability using single year of age because if the solution variability on the 38-bin age schema is zero, then the solution variability for the exact-age schema is restricted to the range of the very narrow bins (except for 85+) in the 38-bin age coarsening. All the other variables (block, sex, race, and ethnicity) would still have zero solution variability. 

The schema used in block-level data has 9,576 possible combinations of sex (2) $\times$ age bins (38) $\times$ race (63) $\times$ ethnicity (2). The sex, age, race, and ethnicity bins used in this schema are those supported by tables P1, P6, P7, P8, P9, P10, P11, P12, P12A-I, and P14 in the 2010 Census Summary File 1 (SF1) \cite{sf1:2010:tech}.
The schema used in the tract and block-level data has 2,884 possible combinations of sex (2) $\times$ age bins (103) $\times$ race (7) $\times$ ethnicity (2) at the tract level and the same 9,576 possible combinations at the block level. Therefore, the full schema supported in tract-level data has 25,956 possible combinations of sex (2) $\times$ age bins (103) $\times$ race (63) $\times$ ethnicity (2). The sex, age, race, and ethnicity bins used in the combined schema are those supported by the list above plus tract-level data in PCT12 and PCT12A-O  \cite{sf1:2010:tech}. Tract-level tables of sex $\times$ age $\times$ race $\times$ ethnicity are less granular for race and ethnicity compared with block-level tables of the same variables; whereas, block-level tables are less granular for age. Combining the tract and block-level tables into the set of constraints used in reconstruction allows results to be computed using both the block-level schema and the more detailed full tract and block-level schema. However, regardless of the schema used to report the results, exactly the same schema was imposed on target linking files when doing reidentification record linkage. Accuracy results reported in the sources cited in this section show the reidenfication based on linking using the block-level schema only was highly successful \cite[Tables 2 and 3]{abowd:hawes:2023} We now relate those results to Muralidhar's critique.

Muralidhar's criteria for evaluating the reconstruction is to assume that only full precision of the age variable (to single year of age at the block level) constitutes success, even when that level of age precision is not necessary to isolate small populations and accurately link reconstructed records to an external data source with names. Muralidhar also makes two significant errors in describing the reconstruction schema: first, he asserts that ``At the block level, the Age variable is grouped (except for ages 20 and 21),'' but SF1 Table P14 also publishes single-year-of-age counts at the block level for ages 0, 1, 2, $\dots$, 19, so that single-year-of-age tabulations are published at the block level up to age 21. Second, Muralidhar describes only 7 levels of race, and treats persons reporting more than one race as a single category, but SF1 Tables P8, P9, P10, and P11 publish counts for all 63 possible (non-empty) combinations of self-reported race interacted with both values of Hispanic or Latino ethnicity. These discrepancies are important, because excluding this extra granularity makes the schema seem artificially coarse, which reduces the extent to which it isolates individual records as unique or nearly unique. Using an incorrect schema also affects the number and character of distinct solutions, although the relationship here is more complicated: for a fixed total population, larger schemas tend to feature more counts that are exactly $0$, which reduces the number of distinct solutions, but larger schemas also increase the number of variables in the linear relaxation of the integer program, which can cause more solutions to exist. A principled approach is required to properly account for these two effects. 

Despite incorrect use of an overly simple schema, Muralidhar's focus on the variability in feasible solutions to the microdata reconstruction problem is an important aspect of bounding the extent to which aggregation alone can provide protection against privacy attacks. The existence of multiple, differing, feasible reconstructions can increase an attacker's uncertainty that they have correctly reidentified a particular data subject. But the Census Bureau's upper bound of zero solution variability for two-thirds of all blocks means that a very large component of the reconstructed data have no solution variability at all.

Muralidhar's use of an incorrect schema leads him to make other significant errors. He claims, for example, that there is no solution variability at the tract level and above. Since there are no tract-level tables that publish the full schema possible when combining tract and block-level tables, as discussed above, solution variability can remain in reconstructions based only on tract-level tables but using the full schema. In unpublished calculations of solution variability at the tract-level for a limited number of tracts, it appears that some tracts do have positive solution variability, although the focus of the study for the full U.S. population was block-level schema solution variability.

At the block level, after the schema used by Muralihdar is corrected to include all the 9,756 possible combinations, there are still sometimes multiple distinct feasible reconstructions, but in most blocks, there is provably a single feasible reconstruction. But since two-thirds of all census blocks have no solution variability, and since the average inhabited 2010 Census block contained 50 persons, it should not be surprising that most blocks in the 2010 Census have substantially less solution variability than the very large block that Muralidhar singled-out in his illustration (block 4000 in tract 5.01 of Laramie County, Wyoming had 450 people) \cite{data:census:gov:url}. For 935 tracts, every single block within the tract shows zero solution variability.

Quantifying solution variability does not require access to any confidential data for verification. The attacker would know that they have the only possible set of reconstructed microdata for such blocks. The attacker would know that population unique persons in the blocks with no solution variability are population unique persons in the confidential data, an inherently disclosive attribute \cite[pp. 42-43]{duncan:et:al:2011:statistical}. Because it is public information that the tables used in the reconstruction were calculated from the Census Bureau's internal swapped file (Hundred-percent Detail File), the attacker would know that they had an exact image of all records in that file, for that block, and for the schema in use. That is, the attacker would know that the publication of separate, low-dimensional tables offered no additional confidentiality protection. It is as if a single table with the full cross-tabulation of variables in the schema had been published. Only homogeneity of large populations and swapping would remain in effect as protections, and, for small populations, the attacker would know that only swapping could possibly be claimed as a source of disclosure protection.

\emph{No reidentification study is required to establish that the SF1 tables once converted to microdata violate the disclosure avoidance rules that were in place for the 2010 Census.} Reidentification studies only confirm that the Census Bureau was correct in holding the official public-use microdata sample to tighter disclosure rules than  swapping, and group quarters synthetic data, age-coarsening, and aggregation used for the tabular summaries. The problem is that too many exact tabular summaries were released, which enabled accurate microdata reconstruction, rendering the release of tabular summaries substantially equivalent to the release of 100\% of the microdata for public-use and undermining the distinction between disclosure rules for these two kinds of products. Accurate reconstruction, in turn, enabled reidentification of very small populations, including population uniques, about which inference is clearly not generalizable, but specific to the person who is population unique.

One reason that at least two-thirds of blocks have no solution variability in the Census Bureau's reconstruction is the extreme granularity of the published 2010 Census data products and the inclusion of block-level invariants for total population and voting-age population. In particular, the block-level population invariant ensures that there is no solution variability on the block identifier for 100\% of the blocks regardless of the set of tract and block-level tables used for the reconstruction. Similarly, the block-level voting-age population invariant ensures that there is no solution variability on whether the individual is an adult or minor for 100\% of the blocks. 

That at least two-thirds of blocks have no solution variability also refutes Muralidhar's concluding exhortation that ``[i]t is up to the Census Bureau researchers to show either that the reconstruction was unique or, at the very least, that most reconstructions lead to the same conclusion regarding the identity of a respondent. Without such proof, claims of confirmed reidentification are highly suspect and easily refuted'' \cite{Muralidhar:2022}. Successful reidentification does not require that the entire reconstruction be unique, although low solution variability can be a dangerous supplement to an existing reconstruction-abetted reidentification attack by enabling an attacker to identify regions where they can remove an important source of uncertainty. Moreover, the Census Bureau would not need to demonstrate low solution variability. Any capable attacker would be able to determine it for themselves, including identification of the blocks with provably zero solution variability in the block-level schema. 

\subsection{Methodological and conceptual errors in papers by Kenny et al}
\label{sec:kenny_errors}

Kenny et al. \cite{kenny:et:al:2021,kenny:et:al:2022} evaluated the impact of demonstration data products released by the Census Bureau computed on 2010 Census data using differentially private mechanisms on redistricting outcomes. They further attempted to measure the disclosiveness of those publications. 

Specifically, using the data demonstration products derived from 2010 Census data, they evaluated their ability to predict race and ethnicity from voter registration and census data. The ground truth for evaluation of prediction consisted of North Carolina voter registration records from 2021 that contained surnames, block, and race/ethnicity data. 

\subsubsection{Their methodology cannot measure disclosure risk}

Kenny et al. cannot, by design, measure disclosure. The meaning of ``disclosure'' is revealing information about an individual that was provided to the agency in a confidential response. If a dataset of records from North Carolina in 2010 enables accurate inference about an individual whose information was not part of the 2010 North Carolina Summary File 1 (e.g., a 2021 North Carolina resident who lived elsewhere in 2010), then that information generally cannot be considered disclosed by the agency (exceptions exist, such as with genetic data of relatives, but such data were not part of the census).

The fact that high quality inference is possible about many North Carolina residents who lived elsewhere in 2010 means that the information they supplied in 2010 (if any) was not used in making this prediction. Hence it was not disclosed. In other words, the prediction of a person's race was based purely on data from \emph{other} people.

For statistical models, the ability to use other people's information to make inferences can be evaluated using leave-one-out cross validation inference because, by definition, the person being predicted is not in the training data. To measure the extra risk of being in the training data, one should compare the in-sample prediction accuracy (accuracy on training data) to the leave-one-out cross-validation accuracy. 

Leave-one-out statistical inferences are the method taken from robust statistics that best captures the formalism required to use the counterfactual posterior-to-posterior comparisons rather than prior-to-posterior comparisons to distinguish a generalizable scientific inference from a privacy-violating inference, as first noted in \cite{dwork:lei:2009}. In leave-one-out inference, the counterfactual world is the one in which the target of an attack has been left out of the statistical model used for inference about or prediction of that person's data. 
Basically, in statistics an inference is not robust if it depends too much on the influence of one or a few observations. The simplest, but usually most computationally intensive way to do this, is to fit the model without using the first observation, then measure the error predicting that observation. This ``leaves out'' observation 1. Then, replace observation 1, leave out observation 2, refit the model, and measure the error predicting observation 2. Repeat until each observation has been predicted from a model where it was left out of the estimation. The cumulative error from such an algorithm estimates the leave-one-out error of the model, also called the out-of-sample prediction accuracy of the model. If the cumulative leave-one-out error of the model is much larger than the cumulative error generated when all observations are used, the inferences from the model are very dependent on one, or a few, observations. Put differently, if the out-of-sample prediction accuracy is much worse than the in-sample prediction accuracy, then the model is not robust---it depends on one, or a few, influential observations. Proper assessment of whether an inference is scientific and generalizable or privacy violating is precisely the assessment of whether the inference is robust, in the leave-one-out sense, or not when the target person is left out of the estimation in the counterfactual world. Robust inferences or predictions are generalizable. Non-robust inferences are privacy violating.

\subsubsection{Disclosure limitation should not prevent high-quality generalizable inference}

Although the methodology of Kenny et al. is not measuring disclosure risk, they do raise the interesting question of whether high-quality inference should be prevented. They argue that privacy violations should be measured as the comparison between inference about an individual after a data release to the inference before the data were released. 
This is the prior-to-posterior methodology described in the main text. As we note in the main text, it is often undesirable to have a disclosure limitation system limit prior-to-posterior inferences, because such limitations would also prevent legitimate scientific findings (such as a dataset demonstrating the causal relationship between smoking and cancer, thus causing a change between prior and posterior distributions). Furthermore, except in rare circumstances, limiting the changes between prior and posterior distributions is known to be impossible \cite{Dwork:Naor:2010}. Indeed, the technical justification that Kenny et al. rely on is a result of Gong and Meng \cite{gongmeng} in which such restrictions were in place but not apparent due to notational issues \changetwo{\cite{gongmengpc}}. \changetwo{See this supplement Section 7, which clarifies the relevant theorem}.

Referring back to the analysis of Kenny et al., if one desired to make high-quality race-ethnicity inferences impossible, then block-level race/ethnicity totals would have to be made highly inaccurate, leading to low-quality redistricting data. This results in a ``have your cake and eat it too'' situation because one cannot make the criticism that race-ethnicity prediction is too good and yet require that redistricting results must match the confidential data.

\subsubsection{High-quality generalizable inference is not a disclosure}

Kenny et al. \cite{kenny:et:al:2021,kenny:et:al:2022} also claim that high quality inference through statistical means is the same as disclosure via reconstruction. This conclusion is not warranted. One reason that they are not the same is due to \emph{calibration}--the confidence they assign to predictions, also known as reidentification precision. They report a roughly 90\% accuracy in predicting race and ethnicity. However, with reconstructed data, one can focus on situations where the inference about whether the record is in the confidential data is guaranteed to be 100\% accurate, relative to the underlying confidential data used in tabulation, the HDF, due to low solution variability. What is important is the difference in prediction precision between populations likely to have high precision due to generalizable inference (modal race-ethnicity groups in a census block, for example) and those likely to fail the leave-one-out generalizable inference criterion (nonmodal race-ethnicity groups in a census block, for example). In the former case (modal) a high-quality inference is often a legitimate scientific one, but in the latter case (nonmodal) the high-quality inference is clearly privacy violating because is not possible without using the target individual's actual race-ethnicity data from the census response.
A consequence of this point is that high quality reconstruction allows inferences on how a person differs from statistical population characteristics (e.g., what one would predict based on what one knows about population characteristics) while high quality inference in the sense of Kenny et al. does not have this property. This is a particular concern in disclosure avoidance.

For example, consider a mixed-race household in which everyone shares the same surname (due to name changes after marriage, say) and lives in a block with only one household. Reconstruction would pose a risk if the household considered their mixed-race status to be confidential, while high-accuracy leave-one-out statistical inference would and should fail in this situation (there are no more data to use) but achieve high accuracy elsewhere (in blocks with many mixed-race households, say). In the sensitive population case (e.g., nonmodal persons in a census block), high accuracy reconstruction is a greater threat specifically because it allows better inference about how individuals differ from population characteristics.

Furthermore, reconstruction that has low solution variability means that there is high confidence associated with any \emph{additional} information that could be attached to a given reconstructed record. If additional tables are added to the reconstruction, more information would be revealed about records and household structures, while purely statistical methods would see their accuracy degrade. In other words, inference via reconstruction carries the risk that additional information available only on a specific census response can also be accurately inferred, whereas, when extra information is added to purely statistical inferences, they suffer as the number of attributes they must predict increases; thus, they do not necessarily carry disclosure risk beyond the attributes they currently predict.

\subsubsection{The major real errors in 2020 Census redistricting data}

Kenny et al. claim that the 2020 DAS TopDown Algorithm (TDA) leads to undercounts of minorities. This statement is false as demonstrated by the correctly calculated error metrics in \cite{wright_irimata,tdahdsr,dsm}. It is not percentage error in the size of a minority population that determine political rights, but the magnitude of the minority population as a percentage of the total population in the political entity. Wright and Irimata \cite{wright_irimata} specifically evaluated the reliability of estimates of the largest racial and ethnic populations as a percentage of total population in political entities (census places and minor civil divisions) and statistical entities (census block groups). This ratio is the measure used in redistricting scrutiny under Section 2 of the 1965 Voting Rights Act. They found that ``[e]mpirical results suggest a minimum TOTAL [population] that is between 450 and 499 people in a block group provides reliable characteristics of various demographic groups in a block group based on the TDA. Similarly, a minimum TOTAL [population] that is between 200 and 249 is observed to provide reliable characteristics for places and MCDs minor civil divisions]'' \cite[p. 1]{wright_irimata}. 

Kenny et al. completely ignore the major acknowledged sources of error in the 2020 Census: differential undercounts resulting from operational and coverage errors due to difficulty contacting hard-to-reach populations, complications of the pandemic, and the political atmosphere surrounding the census, among many other causes. The Post-Enumeration Survey for the 2020 Census \cite{pes:2022:demographic, pes:2022:state}, which is designed to quantify these errors, estimated a differential net undercount of -4.99\% $\pm 0.53$ for Hispanic populations \cite[Table 4]{pes:2022:demographic} and -1.92\% $\pm 0.82$ net undercount in the total population of Texas \cite[Appendix Table 3]{pes:2022:state}. These enumeration issues are much larger than any distortions due to disclosure avoidance. Simulation studies based on estimated data from the 2010 Census Coverage Measurement Program (the post-enumeration survey for the 2010 Census), making them comparable to the analyses in Kenny et al., show that the operational and coverage errors are much more important for political entities, like counties, than disclosure avoidance errors and are of the same order of magnitude as disclosure avoidance errors for census blocks \cite{bell:schafer:2021,census:understanding:variability:2022}.

\subsubsection{Statistical validity of percentage changes in small populations}

Kenny et al. \cite{kenny:et:al:2021} also perform incorrect statistical analysis when measuring percentage changes, their statistic of choice for impugning the accuracy of the 2020 Disclosure Avoidance System (DAS). Their calculations always omit categories where there is no change when comparing the published 2010 Census results in Summary File 1 (SF1) to the results based on applying the DAS to the 2010 Census Edited File (CEF, the unswapped version of the confidential data). They always take the published 2010 Census results as the base, meaning when that result is zero, they cannot compute a percentage change. The correct way to do this, as is well known in economics, is to use an arc percentage change. The numerator is the difference between the two numbers, in this case the output of the 2020 DAS applied to the CEF and the original 2010 SF1 statistic. The denominator is the average of the two. When the two numbers are both zero, report ``no change,'' otherwise, report $100 \times$ the ratio of the numerator to the denominator. This measure is bounded by $\pm200$. More importantly, this measure gives credit when the DAS does not change a zero into a positive number (by design the DAS tries to preserve zeros) and symmetrically assesses when the DAS changes a one to a zero or vice versa, which is appropriate because both SF1 and the DAS output have disclosure limitation errors. In the August 25, 2022 Demonstration Data Product, there are 15.1 billion linearly independent statistics in the ensemble of released tables (the vast majority at the block levels). Of those statistics 13.6 billion (90.6\%) are zero in both the DAS output (demonstration data product) and the original SF1 tables. This calculation uses only public data and can be confirmed directly by comparing the SF1 and DAS output tables for the universe of the demonstration data product. Because the percentage change measures that Kenny et al. use cannot properly account for this accuracy, they grossly overestimate the percentage error in small population statistics. To a first approximation, all their reported percentage changes for small populations should be divided by 10.

\section{Reconstruction of microdata from earlier censuses}
\label{sec:si-reconstruction}

At a meeting of demographers who work closely with the Census Bureau, a member of the group made the statement cited in the main text. The meeting was not public, but the statement was recorded in the transcript. We have cited it anonymously for that reason. The editor of this journal has been supplied with a copy of that meeting transcript.


\section{Clarification of Theorem 1.3 in  Gong and Meng \cite{gongmeng}}
\label{sec:si-gong}
\includegraphics[scale=0.75]{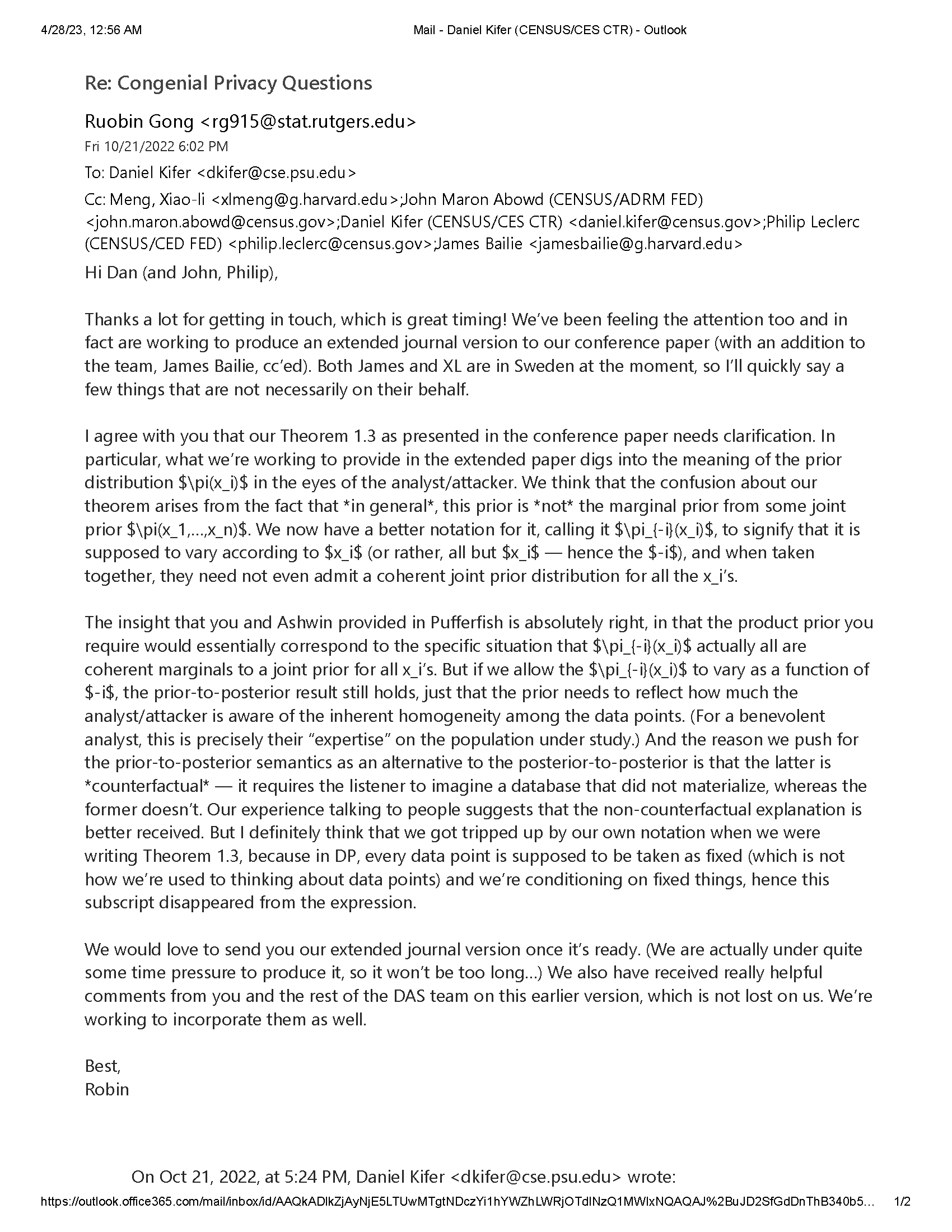}
\newpage
\includegraphics[scale=0.75]{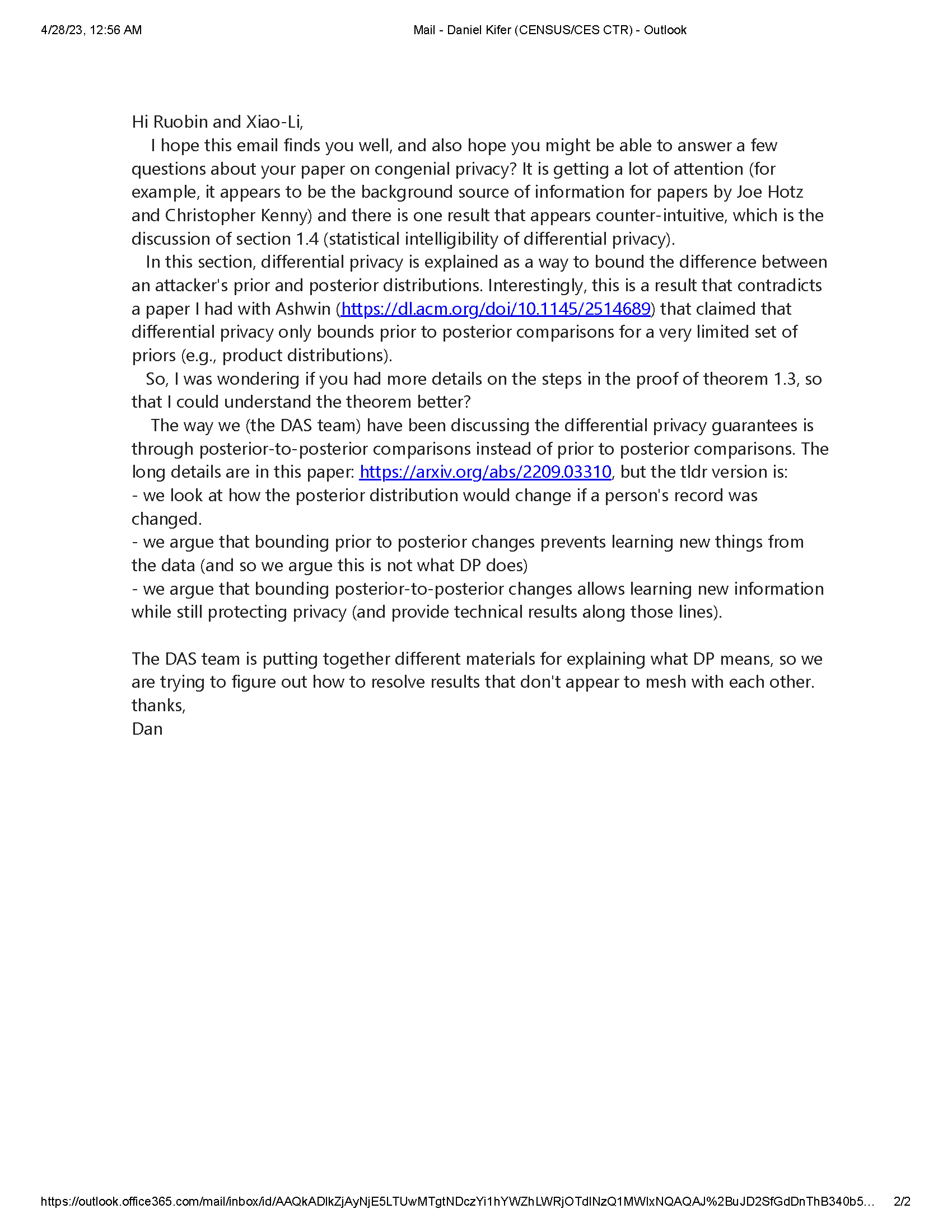}
\newpage



\bibliographystyle{plain}
\bibliography{refs}

\end{document}